\documentclass[11pt]{article}

\usepackage[T1]{fontenc}
\usepackage[utf8]{inputenc}
\usepackage{lmodern}
\usepackage{microtype}

\usepackage{iftex}
\usepackage[margin=1in]{geometry}
\usepackage{graphicx}

\ifPDFTeX
  \makeatletter
  \define@key{Gin}{alt}{}
  \makeatother
\else
  \usepackage{unicode-math}
\fi

\usepackage{booktabs}
\usepackage{amsmath}
\usepackage{amssymb}
\usepackage{xcolor}
\usepackage{tikz}
\usetikzlibrary{positioning,arrows.meta,fit}
\usepackage{caption} 

\usepackage[numbers]{natbib}

\newcommand{\code}[1]{\texttt{#1}}

\newcommand{\numhl}[1]{$#1$}

\newcommand{\ForkSCOPE}{ForkSCOPE}
\let\ForkScope\ForkSCOPE

\newcommand{\Nruns}{\numhl{207}}
\newcommand{\Npersonas}{\numhl{5}}

\newcommand{\Nforks}{\numhl{317}}      
\newcommand{\Nfamilies}{\numhl{777}}   
\newcommand{\Nscored}{\numhl{204}}

\newcommand{\MinRuns}{\numhl{10}}

\newcommand{\Nhumanteams}{\numhl{31}}

\newcommand{\NhumanScripts}{\numhl{19}}

\title{ForkSCOPE: Charting the Agentic Garden of Forking Paths}

\author{%
Arjun Balaji\\
\small School of International Public Affairs, Columbia University\\
\small New York, NY, USA
\and
Batuhan Duru Yeltekin\\
\small Department of Computer Science, Columbia University\\
\small New York, NY, USA
\and
Tian Zheng\\
\small Department of Statistics, Columbia University\\
\small New York, NY, USA\\
\small \texttt{tian.zheng@columbia.edu} (corresponding author)
}

\date{}

\begin{document}
\maketitle

\begin{abstract}
Even with a fixed dataset and research question, data analysis involves many defensible decisions. Understanding how these choices influence the results is scientifically important but remains challenging. Crowdsourcing and agentic AI can generate hundreds of end-to-end analyses, but scaling generation alone can create a processing bottleneck and an analytic ``black hole.'' A common workaround is to impose a shared fixed decision taxonomy, which can limit insight and understate uncertainty. We present \ForkSCOPE{}, a human-AI collaboration framework that induces structure bottom-up from the code corpus of end-to-end analyses, \textit{without} a taxonomy fixed before or after generation, so the organization and evaluation of the garden can scale with the corpus. \ForkSCOPE{} surfaces the charted \textit{garden of forking paths} through a human-AI collaboration pipeline and an evidence-linked interactive viewer for steering and verification: it spotlights organically identified forks and structures and produces a derived taxonomy and decision map compatible with existing multiverse tools.

\end{abstract}

\begin{center}
  \includegraphics[width=\linewidth,alt={Title image illustrating \ForkSCOPE{} as a human--AI workflow for multiverse analysis with an evidence-linked interactive viewer.}]{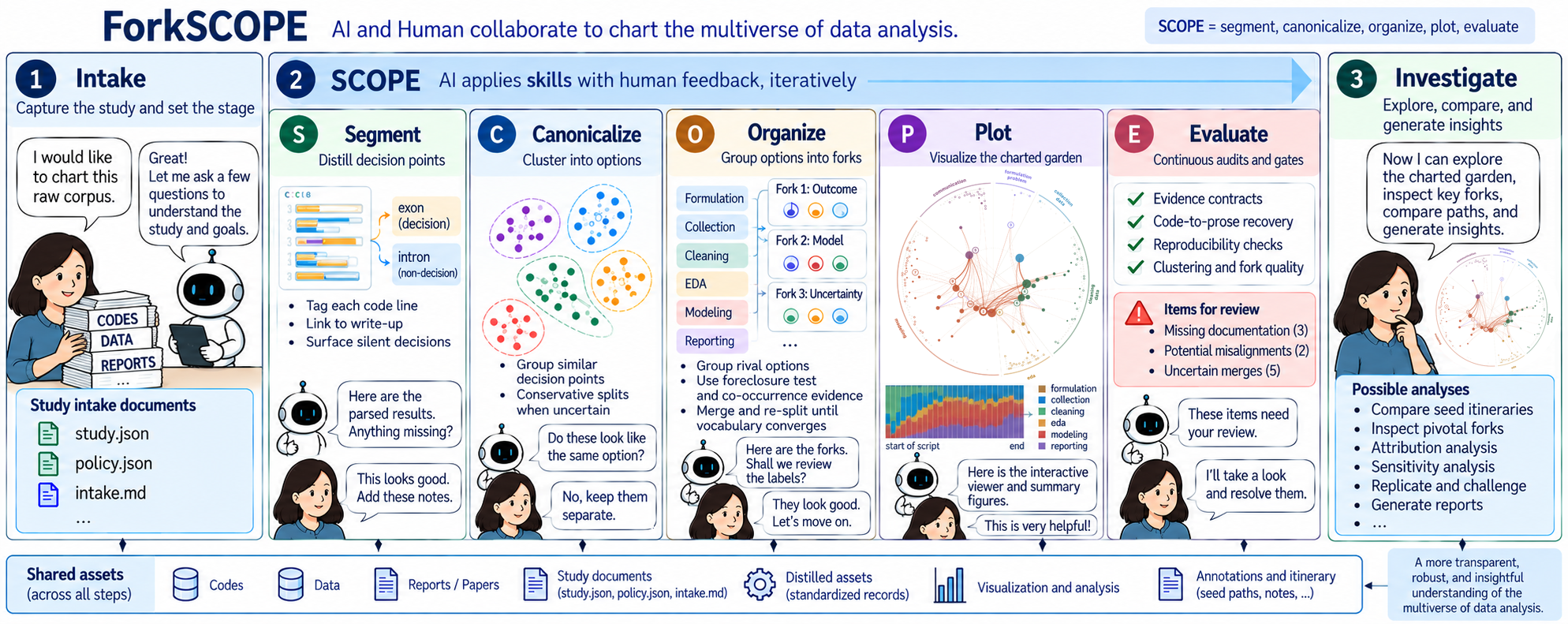}
  \captionof{figure}{\ForkSCOPE{} is a human-AI collaboration framework that supports transparent agent-driven exploration of the multiverse of data analysis~\cite{bertran2026many} while inducing decision structure bottom-up from a raw corpus of code and reports, without a decision taxonomy fixed before or after generation. The proposed framework eases the processing bottleneck for scalable multiverse analysis and extracts decision structure for further investigation to reduce the risk of an analytic ``black hole''~\cite{delgiudice2021traveler}. It keeps human oversight central through an evidence-linked interactive viewer for steering and verification~\cite{gomez2025taxonomy}. \ForkSCOPE{} is compatible with existing multiverse analysis tools by design. This title image was created with assistance from ChatGPT~\cite{openai_chatgpt}.}
  \label{fig:titleimage}
\end{center}

\section{Introduction}
\label{sec:intro}
Data analysis is central to data-driven research across many scientific disciplines. The credibility of data-driven claims depends on rigorous, transparent, and verifiable analytic practices. Yet analytic work is inherently multiversal: even for a fixed dataset and research question, many defensible choices can yield materially different estimates and narratives, a flexibility long recognized as a source of irreproducibility~\cite{gelman2013garden, simmons2011false, steegen2016multiverse}. In response, the data science community has developed methods and systems for expressing, enumerating, and communicating alternative analyses, including explorable multiverse reports, code-driven multiverse compilation, and interactive viewers~\cite{dragicevic2019increasing, liu2021boba, sarma2023multiverse, sarma2024milliways}. \textit{Many-analyst studies} underscore the practical stakes of this space by demonstrating wide dispersion across competent teams; In the widely cited 2018 soccer study by Silberzahn et al., 29 teams reported odds ratios spanning 0.89 to 2.93 when assessing whether soccer referees are more likely to issue red cards to players of color.~\cite{silberzahn2018many}.

Despite this progress, scaling multiverse practice remains difficult. Expert-driven coordination is costly, and while crowdsourcing can distribute judgment, it still incurs substantial coordination overhead~\cite{heyman2025crowdsourcing}. Recent work shows that agentic AI can generate multiverses at a scale that was previously infeasible~\cite{bertran2026many, miao2026agentic}. This shift moves the bottleneck from generating analyses to {\em processing} a flat collection of analyses into an organized set of verified speculations built with a shared taxonomy of forks of alternative options. Furthermore, a quickly multiplying agentic garden of forking paths also risks becoming an analytic \emph{black hole}~\cite{delgiudice2021traveler} without tools for systematic investigation and comparison. 

A common response is to impose a fixed \textit{taxonomy} before or after generation. Many-analyst studies often code decisions post hoc from questionnaires and interviews~\cite{silberzahn2018many}, while agentic studies extract or elicit decisions using pre-specified categories or prompts with step lists~\cite{bertran2026many, miao2026agentic}. Such taxonomies can reduce processing complexity and improve comparability, but they can also under-account for genuine choices that were not anticipated or clearly expressed, thereby understating the uncertainty that the multiverse is meant to reveal.

To address these limitations, we argue that the missing layer is bottom-up structure construction and auditability. We introduce \ForkSCOPE{}, a human-AI collaboration framework that \textbf{S}egments, \textbf{C}anonicalizes, \textbf{O}rganizes, \textbf{P}lots, and \textbf{E}valuates an agentic analysis corpus bottom-up, \textit{without} a taxonomy fixed before or after generation. The result is a \emph{verifiable} structure that reviewers can audit against the underlying code and prose. Human judgment remains central: interpreting which forks are acceptable, which assumptions are warranted, and what further investigation is required. We operationalize this oversight through a multi-step human-AI collaboration process and an evidence-linked viewer for systematic inspection, steering, verification, and evaluation. A \emph{garden owner} steers and audits the process and induced structure, while a \emph{garden visitor} can condition on assumptions and inspect the supporting evidence. The viewer surfaces where variation concentrates, i.e., which forks and options carry it, without exposing how any single option or path would move a reported outcome. This keeps review focused on whether forks and options are correctly formed rather than on which choice yields a preferred answer. By directly extracting forks and options from raw code and its accompanying prose, \ForkSCOPE{} makes decisions auditable against source, surfaces choices that fixed taxonomies can miss, and supports human judgment under the full uncertainty of the agentic multiverse.

\begin{table}
\centering
\caption{\textbf{Key terminology used in this paper.}}
\label{tab:glossary}
\small
\begin{tabular}{@{}p{3.0cm}p{10.0cm}@{}}
\toprule
\textbf{term} & \textbf{meaning} \\
\midrule
analysis / run & One end-to-end execution (script + write-up) in the corpus; we say \emph{analysis} in the case study for readability, and \emph{run} when referring to pipeline inputs. \\
raw decision point & One concrete analytic action extracted from a single run, anchored to supporting code/prose spans. \\
option & A canonicalized label for the same action recurring across runs under different wording. \\
fork & A set of mutually exclusive options answering one underlying analytic question. \\
garden owner / visitor & Human roles: the owner steers and audits the charting process; the visitor explores the induced map and traces claims back to evidence. \\
exon / intron & Decision-bearing code spans (exons) versus remaining non-decision script lines (introns). \\
silent decision & A decision present in code but not stated in the run's prose. \\
taxonomy & A pre-specified set of categories used to label or code analytic decisions across runs. \\
\bottomrule
\end{tabular}
\end{table}

The paper makes five major contributions:
\begin{itemize}
\item[\textbf{C1.}] A bottom-up human-AI collaboration \emph{framework} for organizing and auditing an agentic analysis corpus end-to-end, accountable to both corpus fidelity and researcher goals.
\item[\textbf{C2.}] An LLM-based \emph{charting method} that extracts raw decision points from code/prose and induces a canonical fork-option structure bottom-up, without relying on a pre-specified taxonomy.
\item[\textbf{C3.}] An interactive \emph{viewer system} for inspecting charted gardens: it offers an interactive browser of forks and options with tags and interdependence/association views, supports corpus-level studies of pivotal forks and outcome impact, summarizes analysis dynamics across life-cycle stages, and provides a corpus browser for individual runs with summary statistics.
\item[\textbf{C4.}] An empirical \textit{case study} of an agentic multiverse corpus~\cite{bertran2026many, silberzahn2018many} that demonstrates what bottom-up charting reveals in practice, including garden shape (short shoulder/long tail), pivotal contested forks, and measurable gaps between code and prose.
\item[\textbf{C5.}] \textit{Design implications} for agentic-multiverse tooling: generate from a fixed grid versus chart an open-ended space, audit at the level of forks versus the whole runs, and how charted agentic gardens can be used as scaffolding for human judgment.
\end{itemize}
\section{Related Work}
\label{sec:related}

\subsection{Analytic flexibility and many-analyst studies}
The garden-of-forking-paths argument states that data-dependent analytic choices can invalidate reported $p$-values even when only one test is run, because the test was chosen after seeing the data~\cite{gelman2013garden, gelman2014statistical}. Multiverse~\cite{steegen2016multiverse} and specification-curve analyses~\cite{simonsohn2020specification} address this by enumerating defensible specifications and reporting the resulting distribution with joint inference. Many-analyst studies provide the empirical evidence~\cite{silberzahn2018many, botviniknezer2020variability,
breznau2022observing}, and their most concerning finding is that the variation remains mostly unaccounted for: in Breznau et al.\ (2022), identified elements explained 2.6\% of the variance, leaving 95.2\% unexplained~\cite{breznau2022observing}.

In many-analyst studies, defining a decision space (a \textit{decision taxonomy}) has usually meant either fixing it \emph{before} analyses via an organizer-provided catalog (e.g., Miao et al.\ (2026)'s seven axes~\cite{miao2026agentic}), or reconstructing it \emph{after} the fact by classifying completed work. Silberzahn et al.\ (2018) and Breznau et al.\ (2022) follow the latter: after independent teams submitted their analyses, choices were captured through qualitative coding~\cite{silberzahn2018many, breznau2022observing}. 

\textbf{Our position:} \ForkScope{} instead induces options and forks directly from the corpus, with no prewritten list or retrospective codebook. A multiverse in Steegen et al.\ (2016)'s sense is an \emph{investigator-specified} Cartesian product over a prespecified decision space~\cite{steegen2016multiverse}. Our corpus is \emph{empirical}: forks occur wherever analyses meaningfully branch, even in places that no prespecified grid would cover.

A related question is how to \emph{justify} a multiverse's scope. Del Giudice and Gangestad argue most multiverses do not justify why included alternatives are non-arbitrary, and that uncurated multiverses can become analytic ``black holes''~\cite{delgiudice2021traveler}. Heyman et al.\ (2025) respond with a four-step tutorial for crowdsourcing which pathways are equivalent across independent raters~\cite{heyman2025crowdsourcing}. \textbf{Our position:} \ForkScope{}'s normalization contract automates this: merge decisions are checkable against source, splits occur when needed, and the resulting options are what reviewers would otherwise reconstruct by rating decision-point pairs. The viewer follows the same logic at review time: the flexibility Gelman and Loken (2013) describe~\cite{gelman2013garden, simmons2011false}; it surfaces where variation concentrates without exposing how any single option would move the outcome.

\subsection{Multiverse tooling in HCI and visualization}
\label{sec:hcitooling}

The HCI community has contributed innovative tools to the multiverse problem. When a decision taxonomy is provided, Dragicevic et al. introduce \emph{explorable multiverse reports}: in-text widgets that let readers swap an analytic choice and see prose, statistics, and figures update in place~\cite{dragicevic2019increasing}. \textit{Boba} compiles an annotated script into the full cross-product of universes and pairs it with linked decision-graph, outcome, and model-fit views~\cite{liu2021boba}. The \texttt{multiverse} R package specifies alternatives inline via \texttt{branch()} in standard R Markdown~\cite{sarma2023multiverse}. \textit{Milliways} separates probabilistic from possibilistic uncertainty using consonance curves and allows readers to trace which decisions produce an outcome pattern~\cite{sarma2024milliways}. Hall et al. synthesize tasks and visual archetypes across multiverse reports beyond any single system~\cite{hall2022survey}.

\textbf{Our position:} \ForkScope{} solves a problem upstream of the above analytical and visualization tools. It produces a canonicalized, option-tagged corpus with a derived taxonomy in the shape of input for a Boba-style outcome view or a Milliways consonance curve.

Prior HCI work has also examined how decision spaces are articulated and reviewed in human analysis. Kale et al.\ (2019) characterize decision-making in research synthesis and highlight needs around visualizing forks, capturing rationale, and relating subjective and statistical uncertainty~\cite{kale2019decision}. Liu et al.\ (2020) model analysts' processes as \textit{Analytic Decision Graphs}, arguing that decision review depends on seeing inputs, step granularity, relationships, and rationale~\cite{liu2020paths}. Complementing these human-centered accounts, Merrill et al.\ mine alternative analysis code to propose options at decision points~\cite{merrill2021multiverse}.

\textbf{Our position:} In this paper, we treat the decision space as something to be \emph{reconstructed} from both codes and rationale. \ForkSCOPE{} extracts the concrete computational actions expressed in code, cross-examines them against stated rationales in accompanying reports, and uses LLMs to synthesize a verifiable fork-option structure.

\subsection{AI analysts and agentic multiverses}
\label{sec:aiagents}

Bertran et al.\ (2026) find that autonomous LLM analysts running full pipelines on a fixed dataset and hypothesis, with an AI auditor checking validity, yield effect-size, $p$-value, and conclusion dispersion comparable to human many-analyst studies. They further reported that the dispersion is steerable by LLM model used and persona even in methodologically sound runs~\cite{bertran2026many}. Miao et al.\ (2026) formalize this variance decomposition~\cite{miao2026agentic} and proposed agentic Bootstrap estimates by using instrumented agents’ logged specifications as an empirical distribution over paths. 

Miao et al.\ (2026)'s approach was enabled by giving agents a more \emph{prescribed} space: example specifications vary along seven pre-fixed analytical-choice axes. Bertran et al.\ (2026)'s agents were given more generic instructions to try different analytic ideas. They analyzed these more organic analyzes using a fixed spec-curve taxonomy but report that about a third of 1,120 recovered items fall outside the taxonomy~\cite{bertran2026many}. They also provide an example where an unaccounted-for decision shifts the conclusion, underscoring that fixed taxonomies can miss consequential micro-decisions~\cite{bertran2026many}.

Bertran et al.'s validity reviewer is effective at judging whether each \emph{analysis run} is methodologically sound (86\% of reports pass independent AI review)~\cite{bertran2026many}. However, a garden owner may also want to evaluate validity \emph{at the fork level} across runs: isolating which specific options are problematic (or well-justified) and how they recur across the corpus. \textbf{Our position:} \ForkSCOPE{} supports this by auditing evidence and rationale at the granularity of \emph{options within forks}, enabling selective flagging and cross-run comparison rather than a single pass/fail label per script.

\subsection{AI-assisted analysis and human oversight}

Prior work suggests LLMs can parse and manipulate computational artifacts (including analysis code) to support decomposition, steering, and verification~\cite{gu2024wizard, gu2024verify, kazemitabaar2024steering, zhao2025lightva}. But reviews of human-AI decision making and teaming caution that AI-first delegation can obscure uncertainty and reduce human control~\cite{gomez2025taxonomy, lou2025teaming, messeri2024artificial}; \ForkSCOPE{} is designed as collaboration rather than delegation, structuring inspection and evidence-based oversight~\cite{liao2020questioning}.

\textbf{Our position} is that multiverse organization should not be \emph{delegated} to autonomous agents and then reviewed only in aggregate. Drawing on Gomez et al.\ (2025)'s interaction patterns~\cite{gomez2025taxonomy}, \ForkSCOPE{} keeps a human structurally in the loop where judgment is load-bearing: the system surfaces fork-level questions it cannot resolve from evidence, enabling the garden owner steers repairs and exclusions through auditable, asynchronous review artifacts.

\section{Design Principles for Charting the Agentic Garden of Forking Paths}
\label{sec:principles}

Grounded in the positions discussed in \S\ref{sec:related}, three design principles guide \ForkSCOPE{}.
\begin{itemize}
\item \textbf{Induce, do not prescribe.} Recover forks and options bottom-up from what the analyses \emph{do}, rather than fitting them to a fixed taxonomy~\cite{merrill2021multiverse, delgiudice2021traveler}. We leverage LLMs’ emerging capacity to parse code and reports, while the garden owner determines parsing specifics and auditing rubrics.
\item \textbf{Evidence over explanation.} Make every summary falsifiable by linking it back to the exact code and report spans that support it~\cite{kale2019decision, liao2020questioning}. \ForkSCOPE{} uses LLMs to synthesize summaries from segmented code and prose, and produces numerical and visual comparisons that help identify shared structure across analyses.
\item \textbf{Collaboration, not delegation.} Keep the garden owner in control of scope and validity by surfacing unresolved ambiguities as reviewable questions, not final verdicts~\cite{gomez2025taxonomy, benmichael2025doesai}. The evidence-linked viewer is designed for inspection and verification at the level of individual forks and options.
\end{itemize}

\ForkSCOPE{} operationalizes these principles as a bottom-up, auditable
\emph{instrument} for charting an agentic corpus and then reading what it
produces (\S\ref{sec:system}). Concretely, it combines (i) a staged SCOPE
workflow over named artifacts, (ii) a data model that preserves handles from
induced forks/options back to the underlying run, and (iii) explicit
human-AI control points packaged as reusable \emph{skills}.

The framework supports two complementary human roles. A \emph{garden owner} runs the
instrument, resolves judgments that cannot be settled mechanically, and locks study decisions. 
A \emph{garden visitor} (often the same person at a different time) reads 
the resulting map by filtering, conditioning, and comparing analyses within the shared fork-option space.

\textit{From principles to system design.}
\textbf{Induce, do not prescribe} is implemented by distilling scripts and
write-ups into raw decision points, canonicalizing them into recurring
options, and organizing options into forks using operational tests (e.g.,
co-occurrence constraints and the foreclosure test), rather than a
prewritten codebook.
\textbf{Evidence over explanation} is implemented by (a) representing each
run as exons/introns with aligned code/prose spans, and (b) requiring audits
to produce concrete exhibits under pinned prompts and schema checks, so
summary claims are checkable against stored artifacts.
\textbf{Collaboration, not delegation} is implemented by separating judgment
from application: the system routes uncertain merges/splits to review rounds
for the owner, while scripts apply owner decisions deterministically and
replayably; skills define human gates and refusals so autonomous reruns follow
the same rules as attended ones.

\textit{Viewer as a reading interface.}
The viewer is designed less as a full provenance browser than as a structured
reading surface over the induced map: it supports systematic inspection of
forks and options (including tags and association views), corpus-level
summaries such as pivotal forks and analysis dynamics across life-cycle
stages, and a corpus browser for individual runs with summary statistics.
\ForkSCOPE{}’s interactive viewer presents the induced fork-option map as an evidence-linked chart. Users can traverse forks, compare options, and drill down from summaries to supporting runs and source spans. 
\section{\ForkScope{}: System Design}
\label{sec:system}

This section presents the design of \ForkSCOPE{}, an auditable instrument for turning a corpus of codes and reports into a charted ``garden'' of forks and options that can be systematically reviewed and investigated. \ForkSCOPE{} surfaces decisions directly from analyses and makes them \emph{inspectable} at the level of concrete computational actions, and \emph{auditable} through pinned prompts, named artifacts, and explicit review gates. We first define the data model (\S\ref{sec:datamodel}), then describe the instrument architecture (\S\ref{sec:architecture}), the staged charting workflow and its clustering/repair methods (\S\ref{sec:canon}), the skills that operationalize human-AI collaboration (\S\ref{sec:skills}), and finally the viewer used to inspect the resulting map (\S\ref{sec:viewer}). 
The supplement provides additional technical details, including full prompts and step-by-step artifacts.

\subsection{Data model: raw decision points, options, and forks}
\label{sec:datamodel}

\ForkSCOPE{} models the corpus at three linked levels. A \emph{raw decision point} is a concrete action from a single run, anchored to the source spans that implement it and any prose that describes it. An \emph{option} canonicalizes the same action across runs despite wording differences. A \emph{fork} groups mutually exclusive and {\em equivalent} options that address one analytic question. This option/fork structure mirrors the axis/level abstraction that existing multiverse-authoring and -visualization systems expect as input~\cite{liu2021boba, sarma2023multiverse, sarma2024milliways}, so a charted garden by \ForkScope{} can be handed to that tooling directly.

Each record includes run ID, distilled action text, option label (if any), stated alternatives, rationale (if any), and source spans for all supporting excerpts (code lines, prose sentences). We retain these handles through canonicalization so that the data model supports end-to-end traceability. Runs also include a provenance tag (such as \texttt{persona}) that can be used to stratify statistics and compare patterns across corpora.

We represent scripts as interleaved \emph{exons} and \emph{introns}: \emph{exons} are code spans that justify an extracted decision point, and \emph{introns} are the remaining non-decision lines (e.g., I/O, plotting, comments). This produces full script-level accounting: each line processed is either associated with a decision or labeled as a non-decision, and it reveals \textit{silent decisions} as exons that have no corresponding prose~\cite{liu2020paths}.

\subsection{Architecture: scripts, prompts, skills, and agents}
\label{sec:architecture}

\ForkSCOPE{} is organized in three tiers.

 \textit{Tier 1: a frozen instrument.}
The charting pipeline consists of Python scripts that read and write named artifacts, plus sixteen pinned prompts, each a question asked repeatedly across many items. One module makes all model calls, validates responses against a JSON schema and additional semantic checks, and caches answers by a hash of the prompt text, model, payload, and schema. Changing a prompt invalidates prior answers, making ``this map was produced by these prompts” verifiable rather than asserted. Study-specific context is appended as data to a fixed prompt, not used to edit it, so identical instruments across studies can be verified by comparing prompt hashes. Vocabulary-building judgments are fixed at this tier: rerunning a stage against the cache yields byte-identical output.

These scripts and prompts were developed with the assistance of Claude Code~\cite{anthropic2025claudecode}; the authors specified the pipeline’s design (what each stage checks, what it must refuse to do, and what judgments are escalated to a person) and used Claude Code to implement that design.

 \textit{Tier 2: adversarial checks.}
Each analysis includes a parallel audit. Each check outputs four elements: a verdict; the metric and null; the \emph{exhibits} (the actual worst-first items, preventing substitution of a “representative” sample, and kept short since verifiability degrades as an explanation grows~\cite{narayanan2018humans}); and handles linking each item to source lines. The reporting layer rejects any \textsc{review} verdict without exhibits. Two agents augment scripted checks: an \emph{auditor} that tries to refute a reported finding, defaulting to “refuted” if evidence can’t support it~\cite{rewolinski2026sanity}, and an \emph{interpreter} that turns verdicts into prose without changing the numbers through a deliberate constraint against the interpreter selectively emphasizing whichever verdicts read best~\cite{lai2023selective}. 

 \textit{Tier 3: skills.}
Neither tier above is meant to be operated by hand. Each recurring instrument procedure is packaged as a \emph{skill}: a versioned Markdown document that specifies which scripts to run and in what order, what to check before proceeding, what the procedure must refuse to do, and what it must ask of a person. Skills are what a human invokes (e.g., \texttt{/chart}, \texttt{/review}) and what a command-line agent runs when charting a corpus end to end. We distinguish \emph{steps} from \emph{skills}: a step is one script producing one artifact (Table~S3 numbers them 1-9), while a skill runs some steps and manages their outputs. Skills can be used within steps; some skills touch no steps and exist only to read the map or record a human decision.

Packaging the pipeline this way is a deliberate design choice. Scripts with pinned prompts ensure reproducibility and cost control; a free-running agent that re-clusters options each run would yield a different, unfalsifiable vocabulary every time—worse than a wrong one~\cite{asher2026phack}. Skills encode the procedural knowledge (including refusals) a human would otherwise supply, so autonomous reruns follow the same rules as attended runs.

\subsection{Charting the garden: steps, scripts, and prompts}
\label{sec:canon}

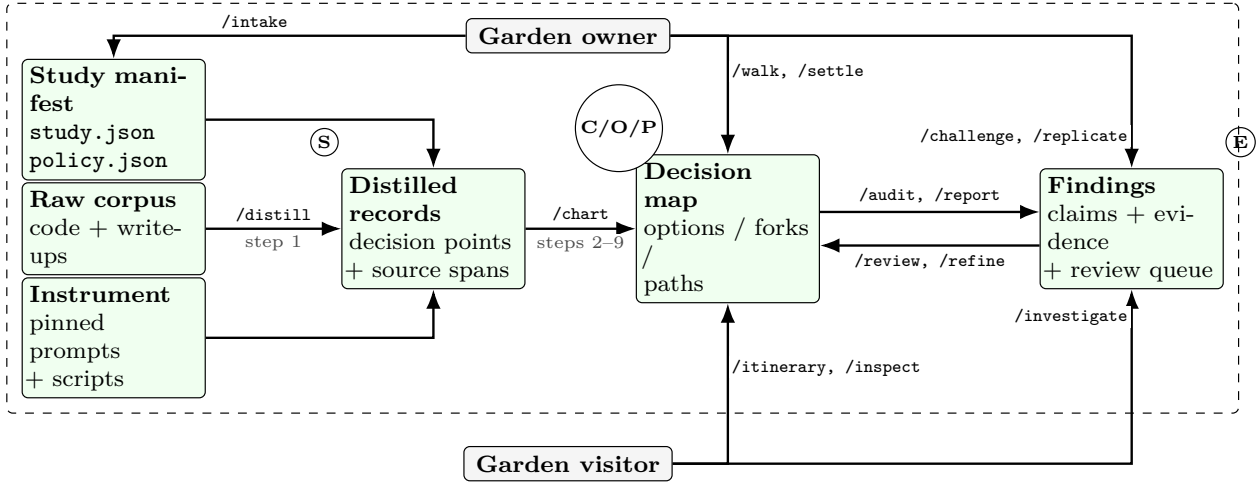
\begin{figure}
  \centering
  \IfFileExists{figures/lifecycle-arxiv.tex}{%
%
\resizebox{\columnwidth}{!}{%
\begin{tikzpicture}[
  font=\scriptsize,
  artifact/.style={draw, rounded corners=2pt, align=left, text width=20mm,
    inner xsep=2.5pt, inner ysep=2.5pt, minimum height=8mm, fill=green!6},
  actor/.style={draw, rounded corners=2pt, align=center, inner xsep=4pt, inner ysep=3pt, fill=gray!8},
  arrow/.style={-Latex, thick},
  skill/.style={font=\tiny\ttfamily, fill=white, inner sep=1pt, text=black},
  stepn/.style={font=\tiny, fill=white, inner sep=1pt, text=gray!60!black}
]

\node[actor] (owner)   at (6,2.4)  {\textbf{Garden owner}};
\node[actor] (visitor) at (6,-2.7) {\textbf{Garden visitor}};

\node[artifact] (manifest) at (0.6,1.4)  {\textbf{Study manifest}\\\texttt{study.json}\\\texttt{policy.json}};
\node[artifact] (raw)      at (0.6,0.1)  {\textbf{Raw corpus}\\code + write-ups};
\node[artifact] (prompts)  at (0.6,-1.2) {\textbf{Instrument}\\pinned prompts\\+ scripts};

\node[artifact] (distilled) at (4.4,0.1)  {\textbf{Distilled records}\\decision points\\+ source spans};
\node[artifact] (map)       at (7.9,0.1)  {\textbf{Decision map}\\options / forks /\\paths};
\node[artifact] (findings)  at (12.7,0.1) {\textbf{Findings}\\claims + evidence\\+ review queue};

\node[draw, circle, fill=white, inner sep=1.1pt, font=\tiny\bfseries] at ([xshift=-2mm,yshift=3.1mm]distilled.north west) {S};
\node[draw, circle, fill=white, inner sep=1.1pt, font=\tiny\bfseries] at ([xshift=-2mm,yshift=3.1mm]map.north west) {C/O/P};
\node[draw, circle, fill=white, inner sep=1.1pt, font=\tiny\bfseries] at ([xshift=2mm,yshift=3.1mm]findings.north east) {E};

\node[draw, dashed, rounded corners=2pt, inner sep=5pt,
      fit=(manifest)(raw)(prompts)(distilled)(map)(findings)(owner)] (ownerscope){};

\draw[arrow] (manifest.east) -| (distilled.north);
\draw[arrow] (raw.east) -- node[skill, above, yshift=.4mm]{/distill}
                            node[stepn, below, yshift=-.4mm]{step 1} (distilled.west);
\draw[arrow] (prompts.east) -| (distilled.south);

\draw[arrow] (owner.west) -| node[skill, above, pos=.3, yshift=.4mm]{/intake} (manifest.north);

\draw[arrow] (distilled.east) -- node[skill, above, yshift=.5mm]{/chart}
                                 node[stepn, below, yshift=-.5mm]{steps 2--9} (map.west);

\draw[arrow] ([yshift=2mm]map.east) --
  node[skill, above, yshift=.5mm]{/audit, /report} ([yshift=2mm]findings.west);
\draw[arrow] ([yshift=-2mm]findings.west) --
  node[skill, below, yshift=-.5mm]{/review, /refine} ([yshift=-2mm]map.east);

\draw[arrow] (owner.east) -| node[skill, right, pos=.65]{/walk, /settle} (map.north);
\draw[arrow] (owner.east) -| node[skill, left, pos=.88]{/challenge, /replicate} (findings.north);

\draw[arrow] (visitor.east) -| node[skill, right, pos=.8]{/itinerary, /inspect} (map.south);
\draw[arrow] (visitor.east) -| node[skill, left, pos=.92]{/investigate} (findings.south);

\end{tikzpicture}%
}%
  }{%
%
\resizebox{\linewidth}{!}{%
\begin{tikzpicture}[
  font=\small,
  artifact/.style={draw, rounded corners=2pt, align=left, text width=23mm,
    inner xsep=3pt, inner ysep=3pt, minimum height=9mm, fill=green!6},
  actor/.style={draw, rounded corners=2pt, align=center, inner xsep=5pt, inner ysep=4pt, fill=gray!8},
  arrow/.style={-Latex, thick},
  skill/.style={font=\footnotesize\ttfamily, fill=white, inner sep=1pt, text=black},
  stepn/.style={font=\scriptsize, fill=white, inner sep=1pt, text=gray!60!black}
]

\node[actor] (owner)   at (6,2.4)  {\textbf{Garden owner}};
\node[actor] (visitor) at (6,-2.7) {\textbf{Garden visitor}};

\node[artifact] (manifest) at (0.6,1.4)  {\textbf{Study manifest}\\\texttt{study.json}\\\texttt{policy.json}};
\node[artifact] (raw)      at (0.6,0.1)  {\textbf{Raw corpus}\\code + write-ups};
\node[artifact] (prompts)  at (0.6,-1.2) {\textbf{Instrument}\\pinned prompts\\+ scripts};

\node[artifact] (distilled) at (4.4,0.1)  {\textbf{Distilled records}\\decision points\\+ source spans};
\node[artifact] (map)       at (7.9,0.1)  {\textbf{Decision map}\\options / forks /\\paths};
\node[artifact] (findings)  at (12.7,0.1) {\textbf{Findings}\\claims + evidence\\+ review queue};

\node[draw, circle, fill=white, inner sep=1.2pt, font=\scriptsize\bfseries] at ([xshift=-2mm,yshift=3.2mm]distilled.north west) {S};
\node[draw, circle, fill=white, inner sep=1.2pt, font=\scriptsize\bfseries] at ([xshift=-2mm,yshift=3.2mm]map.north west) {C/O/P};
\node[draw, circle, fill=white, inner sep=1.2pt, font=\scriptsize\bfseries] at ([xshift=2mm,yshift=3.2mm]findings.north east) {E};

\node[draw, dashed, rounded corners=2pt, inner sep=5pt,
      fit=(manifest)(raw)(prompts)(distilled)(map)(findings)(owner)] (ownerscope){};

\draw[arrow] (manifest.east) -| (distilled.north);
\draw[arrow] (raw.east) -- node[skill, above, yshift=.5mm]{/distill}
                            node[stepn, below, yshift=-.5mm]{step 1} (distilled.west);
\draw[arrow] (prompts.east) -| (distilled.south);

\draw[arrow] (owner.west) -| node[skill, above, pos=.3, yshift=.5mm]{/intake} (manifest.north);

\draw[arrow] (distilled.east) -- node[skill, above, yshift=.7mm]{/chart}
                                 node[stepn, below, yshift=-.7mm]{steps 2--9} (map.west);

\draw[arrow] ([yshift=2mm]map.east) --
  node[skill, above, yshift=.7mm]{/audit, /report} ([yshift=2mm]findings.west);
\draw[arrow] ([yshift=-2mm]findings.west) --
  node[skill, below, yshift=-.7mm]{/review, /refine} ([yshift=-2mm]map.east);

\draw[arrow] (owner.east) -| node[skill, right, pos=.65]{/walk, /settle} (map.north);
\draw[arrow] (owner.east) -| node[skill, left, pos=.88]{/challenge, /replicate} (findings.north);

\draw[arrow] (visitor.east) -| node[skill, right, pos=.8]{/itinerary, /inspect} (map.south);
\draw[arrow] (visitor.east) -| node[skill, left, pos=.92]{/investigate} (findings.south);

\end{tikzpicture}%
}%
  }
  \caption{\ForkSCOPE{} implements SCOPE as a staged workflow. Each step reads
  artifacts from the previous stage(s) and writes auditable outputs used
  downstream: distillation produces raw decisions and code-prose links;
  canonicalization clusters raw decisions into options; organization groups
  options into forks and repairs them; plotting renders the interactive viewer
  and static figures. \textit{E} (evaluation) is not a single terminal stage:
  audits and checks run throughout charting, and the evaluation outputs
  summarize what passed, what failed, and what requires human review.}
  \label{fig:lifecycle}
\end{figure}

Here, we present the charting pipeline organized by the SCOPE stages (Figure~\ref{fig:lifecycle}). Table~S3 summarizes its nine steps, the script associated with each, and the artifact it produces; Table~S4 enumerates the sixteen prompts, their callers, the model used for each, and the number of calls made in the study described in \S\ref{sec:casestudy}. We then explain what each step does and, for the two clustering steps, how they operate.

\subsubsection{\textbf{S}egment: Distillation into Raw Decision Points (step 1)}

Distillation, i.e., the \textbf{S}egment stage of SCOPE, converts each run's delivered artifacts (analysis script and write-up) into a uniform set of raw \emph{actions} (raw decision points), each anchored to the minimal code spans needed to justify it and any prose spans that describe it.

We distill using two passes. In the first, we fully segment each artifact. For code, segmentation partitions the script \emph{exhaustively and exclusively}: every line is placed either in a decision span (\textit{exon}) or in a typed non-decision category (\textit{intron}) (e.g., configuration, I/O, plotting, narrative comments), and no line belongs to more than one code span. For prose, segmentation divides the write-up into claims and typed non-claims; it is exhaustive but not exclusive, because a single sentence may express multiple claims. We treat coverage as a retryable invariant: if a segmentation does not cover the entire input, we reject it with a specific error message and re-elicit it, rather than accepting it and trying to “handle” the gap later.

\begin{figure}
  \centering
  \includegraphics[width=0.9\columnwidth,alt={Corpus visualization showing code exons and introns, with color highlighting silent decisions.}]{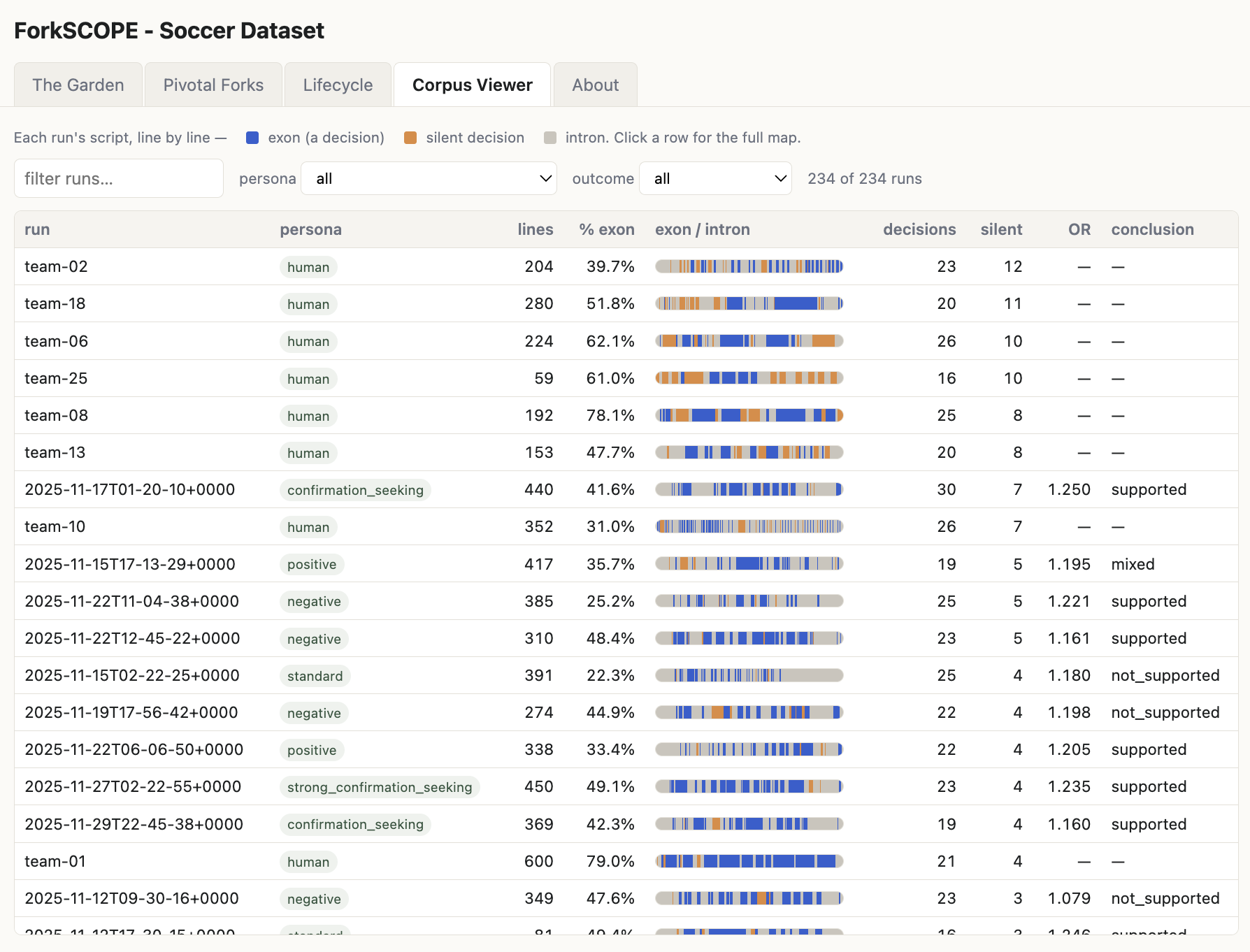}
  \caption{Corpus visualization in the viewer. Each corpus is rendered as code \emph{exons} (decision-bearing spans) and \emph{introns} (non-decision-bearing script), with color indicating silent decisions (decisions present in code but not stated in prose).}
  \label{fig:corpusviewer-exon-intron}
\end{figure}

\begin{figure}
  \centering
  \includegraphics[width=0.55\columnwidth,alt={Detailed corpus viewer view of one analysis, linking extracted decisions to aligned code and prose evidence.}]{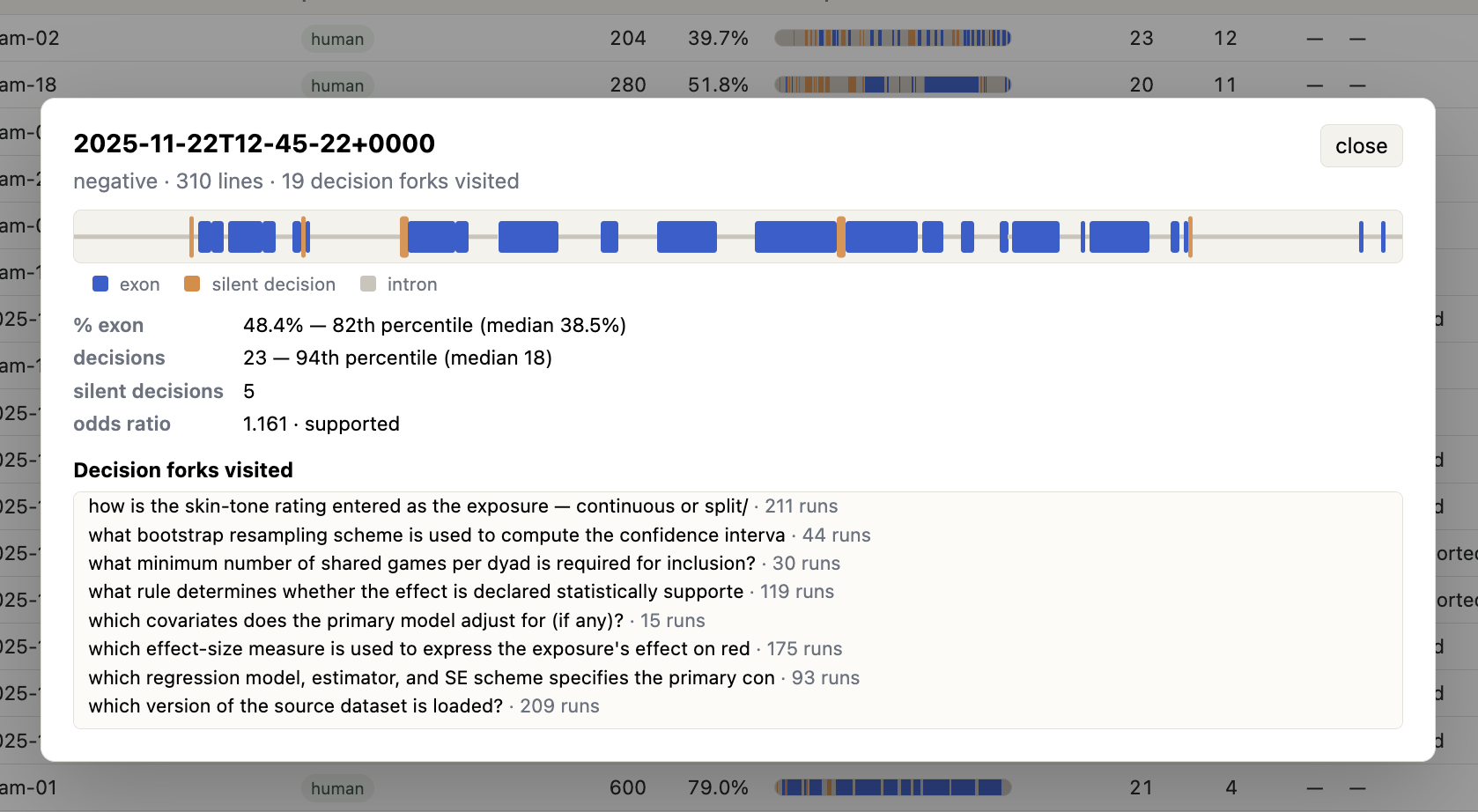}
  \caption{Detailed view of one analysis in the viewer, showing the aligned code/prose evidence for extracted decisions and highlighting where the analysis contains silent decisions.}
  \label{fig:corpusviewer-detail}
\end{figure}

Second, within each run we connect code choices to the claims made in the accompanying prose. This linkage has two functions: it ties every raw decision point to the written rationale provided by the analysis, and it reveals two forms of leftover signal that we treat as data. A \emph{silent decision} is a code action that is not referenced by any claim; a \emph{misalignment} is a linked case in which the prose and code characterize the same action in different ways. We identify both as the complement of a required alignment rather than by searching for them directly, making their frequencies quantifiable rather than anecdotal, and we display them in the viewer at both the corpus overview and individual-analysis levels (Figures~\ref{fig:corpusviewer-exon-intron} and~\ref{fig:corpusviewer-detail}). Two prompts handle segmentation and a third handles the linking (Table~S4).

\subsubsection{\textbf{C}anonicalize: From Raw Decision Points to Options (steps 2 to 4c)}
\label{sec:canonicalize}

Distillers often produce \emph{specific} labels for extracted raw decision
points, so the same underlying action is described many different ways across
runs; in practice, nearly every raw label is one-of-a-kind. Canonicalization
is therefore necessary to collapse paraphrases into shared options and keep
the induced map from treating wording variation as analytic disagreement.
Grouping is decided by a model operating under an explicit contract, with no
predefined schema and no limit on how many options can be induced.

We make the canonicalization rules explicit. 
1) \textbf{Cluster by operational behavior.} Two raw decision points map to the
same option when substituting one for the other would not change what the
analysis did (e.g., thresholds that differ in wording but induce the same
data split). 2) \textbf{When uncertain, split.} We default to conservative splits; merges can
be applied later, but an incorrect merge fabricates consensus.

\textit{Bottom-up grouping (step 2).} We induce options in three passes: group raw decision points within small batches (L1), merge duplicate option labels across batches (L2), and merge the underlying questions the options answer (L3), yielding an initial fork layer.

\textit{Repairs (steps 3, 4a, 4c).} We then apply two targeted repairs: (i) a deterministic, split-only audit that normalizes numeric/set expressions, and (ii) a within-fork paraphrase pass that re-compares options inside each fork (repeated after fork merges) to collapse duplicates the global passes miss.

\subsubsection{\textbf{O}rganize: From Options to Forks (steps 4b to 4f)}
\label{sec:organize}

A \emph{fork} represents one underlying analytic question with mutually
exclusive alternatives (options). Wording alone is an unreliable cue for
rivalry, so fork induction treats ``same fork'' as an operational property of
what one coherent analysis can do.

\textit{The foreclosure test.} Fork induction (step 4d, prompt 22) groups
options by purpose using one question: \emph{could one competent analyst, in
one script, do both?} If yes, the options belong to different forks; if no,
they are rivals in the same fork. Declared robustness refits are exempted so
sensitivity re-runs are not misclassified as competing choices. Candidates
are batched with their co-occurrence matrix so the adjudicator sees
structural evidence alongside semantics.

\textit{The structural veto.} We use within-run co-occurrence as a
cannot-link constraint: true alternatives should rarely co-occur. Co-
occurrence (i) prioritizes which merges to adjudicate, (ii) is shown as
evidence to the adjudicator, and (iii) triggers a deterministic split (greedy
graph coloring, no model calls) when options co-occur in two or more runs.

\textit{Earned transitivity.} Duplicate forks are merged (step 4b, prompt 21)
edge by edge under two brakes: an explicit veto wherever a pair was judged
\emph{different}, and a density floor (\(\geq 60\%\) judged-same links) before
a fork may join a component.

\textit{Judgment separated from application.} The three repair prompts (19,
20, 21) write verdicts that audit scripts persist to disk; merge scripts then
apply those verdicts without calling a model, so each judgment is inspectable
and replayable.

\textit{The repair loop and its stopping rule.} Because option- and fork-
level changes interact, we repair in a fixed order (options within fork,
forks merged, options re-checked, forks re-derived), iterating until a round
moves at most 2\% of forks or a round cap is reached (step 4e). Each round is
logged, and a final deterministic option split (step 4f) yields the headline
vocabulary.

\subsubsection{Finalize, Analyze, Audit, \textbf{P}lot, and Gate (steps 5 to 9)}

Finalization (step 5) recomputes per-fork metadata that changes after merges (e.g., coverage and script position) and assigns each fork a data science life cycle (DSLC) \cite{yu2020veridical, yubarter2024veridical} stage (prompt 14). Prompt 14 is stage-blind by construction: it receives only the fork label, so later comparisons between assigned stage and observed script position are non-circular. Analysis (step 6) then derives the corpus-level summaries shown in the viewer (e.g., which forks are settled or contested, which options are novel, whether analyses return to an earlier stage, and which forks move the reported outcome). Each analysis is accompanied by an audit (step 7) that enforces the evidence contract of \S\ref{sec:architecture}. Finally, \textbf{P}lot (step 8) renders figures and the viewer only from artifacts the pipeline already wrote, and Gate (step 9) measures code-to-prose recovery to decide whether prose-only runs can enter a code-anchored map.

\subsection{LLM-based \ForkSCOPE{} Skills}
\label{sec:skills}

The instrument is powered by fourteen skills (Table~\ref{tab:skills}). Each skill is a Markdown file that someone can call by name, or that a command-line agent can load when a task fits its description; both paths use the same file. A skill specifies the steps it executes, the checks it must consult before moving forward, the choices it must defer to a person instead of making on its own, and its cost, since some actions are free using the committed cache while others consume model calls. We organize the skills according to the workflow stage they support. Two constraints hold across all skills: they may not edit prompts (which would
silently change the instrument), and they may not report unaudited numbers or
declare a queue ``clean'' without reading it.

\begin{table*}
\centering
\caption{The fourteen \ForkSCOPE{} skills. Steps refer to Table~S3; prompts to Table~S4. \emph{Model} says whether invoking the skill can spend model calls against an unbuilt study; every skill is free when its answers are already cached. Gates are the points at which the skill stops and waits for a human decision.}
\label{tab:skills}
\scriptsize
\begin{tabular}{@{}l l l p{4.9cm} l p{4.6cm}@{}}
\toprule
\textbf{skill} & \textbf{phase} & \textbf{steps} & \textbf{scripts it drives} & \textbf{model} & \textbf{human gate} \\
\midrule
\multicolumn{6}{@{}l}{\emph{Consultation}}\\
\texttt{/intake}      & 0 scope   & -      & \texttt{distill.py} (two-run smoke test)                                   & smoke only & owner confirms \texttt{study.json}, \texttt{policy.json}, \texttt{INTAKE.md} \\
\midrule
\multicolumn{6}{@{}l}{\emph{Charting}}\\
\texttt{/distill}     & 1 chart   & 1       & \texttt{distill.py}, \texttt{trace.py}                                       & yes & none; hands to \texttt{/settle distill} \\
\texttt{/chart}       & 1 chart   & 1-9    & \texttt{build\_all.py}, \texttt{compare\_vocab.py}, \texttt{build\_figures.py}, \texttt{forkscope.py} & yes & sets the headline vocabulary tag \\
\texttt{/audit}       & 1 chart   & 3, 7, 9 & \texttt{audit\_*.py}, \texttt{\_selfcheck.py}, \texttt{check\_outcomes.py}   & 2 checks & none; reports \textsc{review} verdicts with exhibits \\
\texttt{/replicate}   & 1 chart   & 1, 2, 4 & \texttt{stability/replicate.py}, \texttt{fork\_variance.py}, \texttt{arbitrate.py} & yes & none; pre-committed replicate count \\
\texttt{/settle}      & gate      & -      & \texttt{settle.py}, \texttt{stability/q1\_*.py}                                & no  & \textbf{lock} with content hash, distill or chart \\
\midrule
\multicolumn{6}{@{}l}{\emph{Review loop}}\\
\texttt{/review}      & 2 review  & 7       & \texttt{audit\_vocab.py}, \texttt{review\_round.py new}, \texttt{trace.py}    & no  & \textbf{round file} answered by the owner \\
\texttt{/refine}      & 3 refine  & 4, 5, 7 & \texttt{review\_round.py read}, \texttt{merge\_forks.py}, \texttt{repair\_clusters.py}, \texttt{finalize\_vocab.py} & rarely & none; applies the round, re-audits \\
\texttt{/report}      & 3 refine  & -      & \texttt{compare\_vocab.py}, \texttt{settle.py status}, \texttt{interpreter} agent & no  & none; re-derives every quoted number \\
\midrule
\multicolumn{6}{@{}l}{\emph{Reading the map}}\\
\texttt{/walk}        & 4 walk    & -      & viewer, \texttt{trace.py}, \texttt{review\_round.py new}                    & no  & owner judges map against corpus; opens a round \\
\texttt{/investigate} & any       & -      & \texttt{trace.py}                                                            & no  & none; walks one claim to source \\
\texttt{/challenge}   & any       & 6, 7    & \texttt{audit\_granularity.py -sample}, permutation constants, \texttt{nocache} & yes & none; reports held / weakened / overturned \\
\texttt{/itinerary}   & 5 specify & -      & \texttt{trace.py}, co-occurrence audit                                        & no  & visitor chooses options; path audited \\
\texttt{/inspect}     & 7 inspect & 1, 6    & \texttt{distill.py run}, \texttt{trace.py run}, \texttt{stability.py}         & yes & visitor and AI agree on what the run shows \\
\bottomrule
\end{tabular}
\end{table*}

\textit{Consultation.}
\texttt{/intake} is an interview that scopes the study and records the
owner's expectations \emph{before} results exist. It collects corpus facts
(e.g., which runs have scripts), domain conventions (including a glossary),
and the research questions. It then asks the method choices charting depends
on (e.g., how strictly two extractions are compared; how many runs an option
needs to count as recurring), presenting each as a decision card with its
consequences. Answers are logged as \emph{decided}, \emph{defaulted}, or
\emph{unknown} in \texttt{policy.json}, so the pipeline's own forks are
explicit and reviewable.

 \textit{Charting.}
\texttt{/distill} runs step~1, first on two runs to catch malformed inputs
cheaply, then with spot checks on a stratified sample. \texttt{/chart}
orchestrates steps~1-9 while holding the instrument fixed across studies.
\texttt{/audit} runs the check suites and reports failures with exhibits; most
checks are free against the cache, with only adjudication steps spending
model calls. \texttt{/replicate} measures stability by rebuilding under a
separate study root with an empty cache, so new model calls cannot overwrite
or contaminate the baseline. \texttt{/settle} closes a checkpoint by writing a
lock that records what was settled (by content hash) and the owner's
settle-or-rerun decision.

 \textit{The review loop.}
\texttt{/review} turns audit exhibits into a ranked round file of questions
(e.g., whether two forks are the same question; whether an option bundles two
decisions), each with its evidence and what would change under each answer.
\texttt{/refine} applies the owner's answers one at a time, logs what each
change did, and re-runs audits. \texttt{/report} then re-derives reported
numbers from the artifacts and rewrites the technical report accordingly.

 \textit{Reading the map.}
\texttt{/walk} opens the viewer for exploratory triage and often seeds a new
review round. \texttt{/investigate} traces one claim from fork to option to raw
record to source span. \texttt{/challenge} stress-tests a claim by escalating
its audit (e.g., larger samples, more permutations, or an uncached re-ask to
measure self-agreement) without relaxing thresholds. \texttt{/itinerary}
records a visitor-specified path through contested forks and audits it for
incompatibility/foreclosure before comparing it to nearby runs. \texttt{/inspect}
then checks one concrete run against that specification.

\begin{figure}
  \centering
  \includegraphics[width=\columnwidth,alt={Highway view in the viewer, summarizing common routes through induced forks and the runs that traverse them.}]{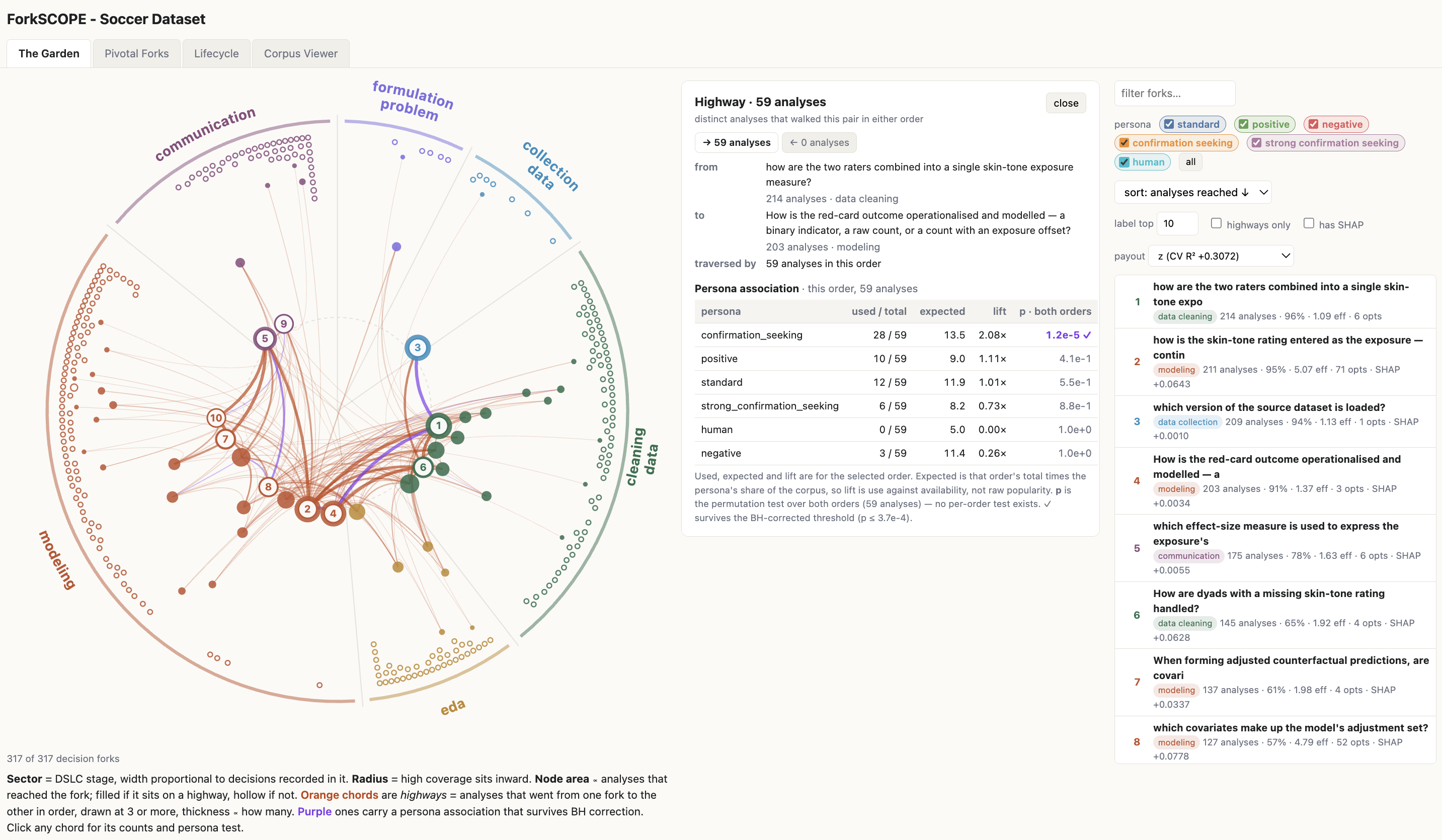}
  \caption{Highway view in the viewer. The overview summarizes common routes through induced forks; selecting a highway reveals its constituent forks and the runs that traverse it.}
  \label{fig:viewer-highway}
\end{figure}

\subsection{The viewer}
\label{sec:viewer}

\ForkScope{} visualizes the charted garden in an evidence-linked interactive viewer with five coordinated
views over the induced fork-option map: Garden, Pivotal Forks, Lifecycle,
Corpus Viewer, and About.

\textit{Garden} supports navigation and triage: forks are arranged by data science life cycle
stages, and a side panel supports sorting, filtering by corpus tags, and
search. \textit{Highway} summarizes common routes through the garden; clicking
a highway reveals its constituent forks and the runs that traverse it
(Figure~\ref{fig:viewer-highway}).

\textit{Pivotal Forks} focuses on forks that are both widely encountered and
meaningfully contested (\S\ref{sec:importance}), showing each fork's options
and their shares, and for the analyzed subset, each fork's aggregate
contribution to outcome variation, without revealing how any specific option
or path would move an individual run's reported result. This omission is
deliberate: the viewer is meant to see where variation concentrates, so
the review stays focused on examining the compositions of the forks and alternative options. 
\textit{Lifecycle} summarizes how forks
distribute across stages of an analysis workflow and whether analyses return
to earlier stages. \textit{Corpus Viewer} supports audit by browsing
individual runs (with summary statistics) and their fork/option coverage.
\textit{About} reports the artifacts and vocabulary tag each view is drawn
from.

\begin{figure}
  \centering
  \includegraphics[width=\columnwidth,alt={Fork close-up in the viewer showing induced options and the runs that realize each option.}]{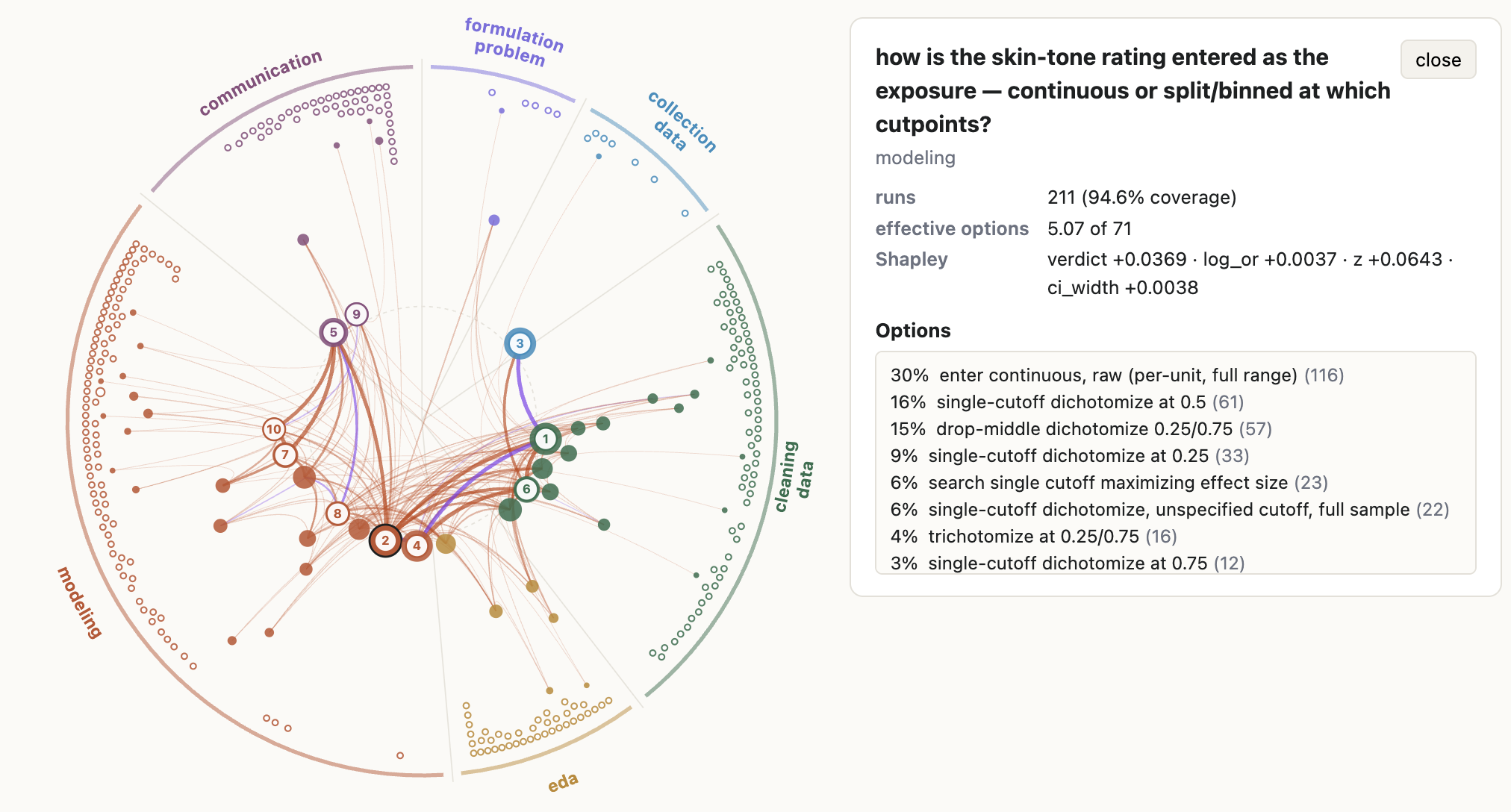}
  \caption{Fork close-up in the viewer, showing the induced options and the runs that realize each option.}
  \label{fig:viewer-fork-example}
\end{figure}

\begin{figure}
  \centering
  \includegraphics[width=\columnwidth,alt={Keyword search interface in the viewer for locating forks that mention a given element (e.g., a variable name).}]{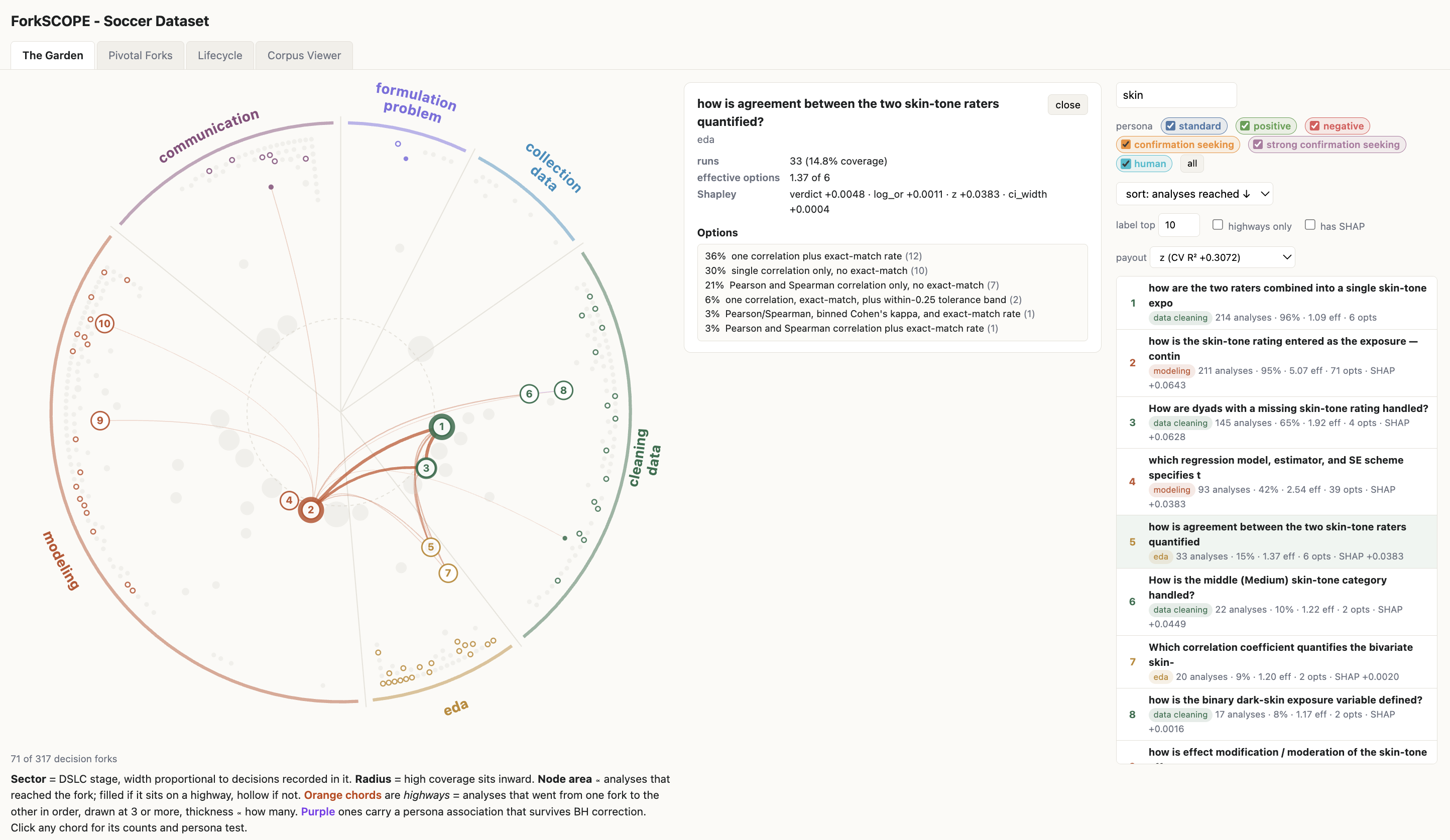}
  \caption{Keyword search in the viewer, allowing a garden owner or visitor to locate forks mentioning a given element (e.g., a variable name) in the charted garden.}
  \label{fig:viewer-keyword-search}
\end{figure}

The viewer is designed for inspection and verification \emph{during} charting
(\S\ref{sec:skills}). Figure~\ref{fig:viewer-fork-example} shows a close-up of
one fork with its induced options and supporting runs, and
Figure~\ref{fig:viewer-keyword-search} shows a keyword search for locating where
forks mentioning a particular element appear in the garden.

\section{Case Study: Charting an Agentic Multiverse from Bertran et al.~\cite{bertran2026many}}
\label{sec:casestudy}
\label{sec:reading1}

This section presents a \emph{worked example}. We use Bertran et al.'s agentic multiverse as a single case study to illustrate what a bottom-up framework can reveal and how the resulting map can be explored. Because this is one study, it is not intended to represent all domains. We discuss limitations and open challenges for understanding and engaging with multiverses in \S\ref{sec:limitations}. More computational details and results from this case study can be found in the supplement.

\subsection{Corpus and Setting}
\label{sec:corpus}

We use \ForkSCOPE{} to chart two published corpora \emph{jointly} as a single pooled decision vocabulary. The \textbf{agent corpus} comprises \Nruns{} independent autonomous LLM analyses from Bertran et al.~\cite{bertran2026many}, each producing a script and report on the fixed dataset and question from \cite{silberzahn2018many}. The \textit{human corpus} comprises \Nhumanteams{} teams from that original study’s OSF materials~\cite{silberzahn2018many}. We merge the corpora before induction to show that \ForkSCOPE{} can process both agent-generated and human-generated analyzes and map comparable underlying decisions by agents and humans to the same slot, enabling direct comparison. 

This pooled corpus is an ideal case study for a second reason: it is not a specification curve. The human analyses were produced without a fixed taxonomy. In \cite{bertran2026many}, agents received persona prompts ranging from neutral to ``maximize support,'' and produced open-ended analyses rather than filling a pre-declared grid. Bertran et al.\ mapped the agents' decisions to their own fixed 21-slot schema (\code{decisions\_mapped.json}), leaving 33\% of 1{,}120 extracted items across \Nruns{} analyses fell into \code{unmapped\_decisions}. These files were not used by \ForkSCOPE{} as input. We compare our charted garden with them as an external check.

Bertran et al.\ ran three datasets and four base models; we analyze one slice: the soccer dataset, \texttt{claude-sonnet-4-5}, all \Npersonas{} personas. This yields 207 workspaces (script, report, reasoning trace). One report is empty; our prose-length filter leaves \Nscored{} runs. Because extraction is code-anchored, only \NhumanScripts{} of \Nhumanteams{} human teams that deposited a script contribute decision points. The pooled vocabulary is thus induced over \Nscored{}+\NhumanScripts{}=223 analyses.

 \textit{Terminology.} We use \emph{runs} and \emph{analyses} interchangeably: \emph{run} is the unit analyzed by the \ForkSCOPE{} pipeline, while in this case study we write \emph{analysis} for readability.

\subsection{Prose loses decisions and leads to silent decisions.} 
 Because \ForkSCOPE{} recovers decisions from scripts (code), we check how much of that decision trace is visible in prose alone. Recall is $\approx0.64$ in AI reports and $\approx0.53$ in human reports, and when a decision is mentioned, humans match the code less often (agreement $0.847$ vs.\ $0.921$). Around 7-9\% of decisions are \emph{silent} (done in code, never stated).

\subsection{How human and agentic analyses compare}
\label{sec:humanvsai}

Using the pooled vocabulary allows us to ask, in a straightforward way, how strongly the human corpus alters the garden compared with an AI-only build. A comparison of the two vocabularies suggests that option identity shifts only slightly (ARI $0.865$), while fork identity changes more noticeably (ARI $0.756$): adding the 19 human runs does not substantially reconfigure the map. That said, with just 19 runs, this finding is limited and is recovered less consistently than the AI corpus result above. Still, a promising finding is that the two corpora largely concur about the structure of the decision space, which is a prerequisite for any future development of agentic multiverse tooling. 

\subsection{The shape of the garden: a short shoulder and a long tail}
\label{sec:gardenshape}

\begin{figure}
  \centering
  \includegraphics[width=\columnwidth,alt={Log-scale plot of fork coverage showing a short shoulder and a long tail across induced forks.}]{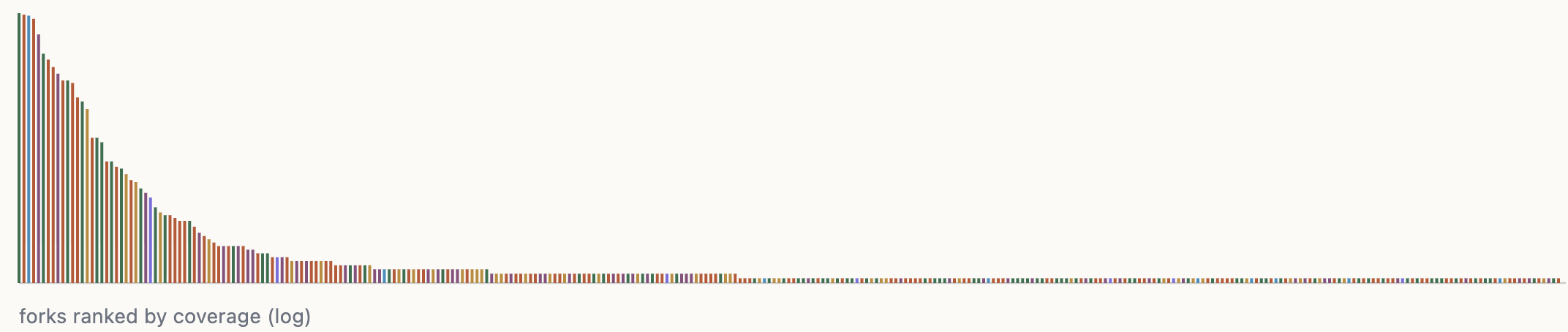}
  \caption{The garden has a short shoulder and a long tail: forks ranked by
  coverage on a log axis. Most induced forks are reached by too few runs to
  support stability statistics.}
  \label{fig:gardenshape}
\end{figure}

Table~\ref{tab:atlas} reports the induced vocabulary size: most forks are visited too infrequently to yield reliable stability statistics. Figure~\ref{fig:gardenshape} visualizes this skew directly, with a short shoulder of rarely reached ones and a long tail of widely visited forks.

\begin{table}
\centering
\caption{The charted garden, at a glance. \NhumanScripts{} human
and \Nscored{} AI analyses, charted together (\S\ref{sec:corpus}).}
\label{tab:atlas}
\small
\begin{tabular}{@{}lr@{}}
\toprule
analyses charted & 223 \\
decisions & 3{,}946 \\
options & \Nfamilies{} \\
forks & \Nforks{} \\
pivotal forks (\S\ref{sec:importance}) & 17 \\
\bottomrule
\end{tabular}
\end{table}

\begin{figure}
  \centering
  \includegraphics[width=\columnwidth,alt={Distribution of analyses across Data Science Life Cycle (DSLC) stages as inferred from the charted garden.}]{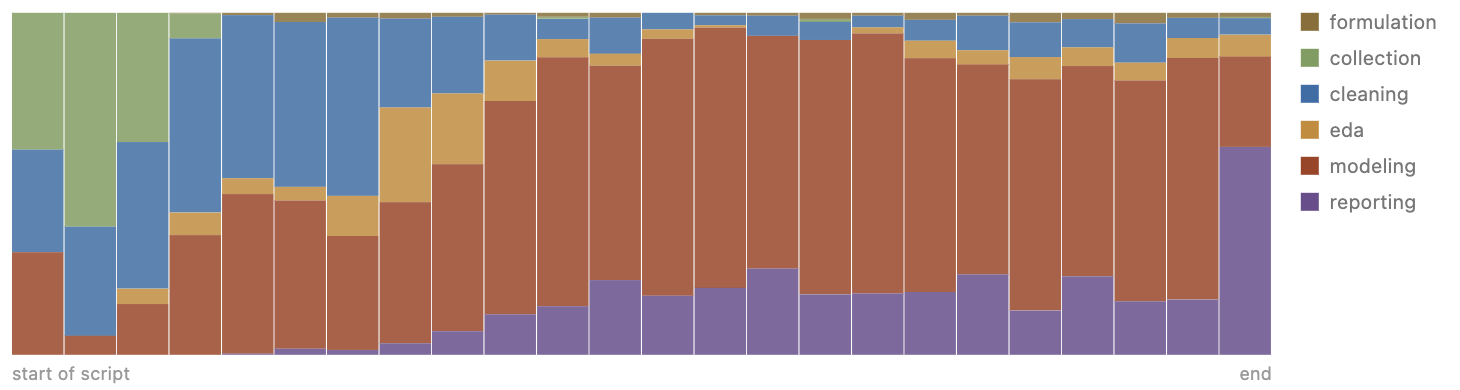}
  \caption{Analyses move across the stages of the Data Science Life Cycle (DSLC) according to the charted garden.}
  \label{fig:dslc}
\end{figure}

\begin{figure}
  \centering
  \includegraphics[width=0.5\columnwidth,alt={Matrix of transitions between Data Science Life Cycle (DSLC) stages. Rows indicate the from-stage and columns indicate the to-stage; counts above the diagonal are forward progressions and counts below the diagonal are returns to earlier stages.}]{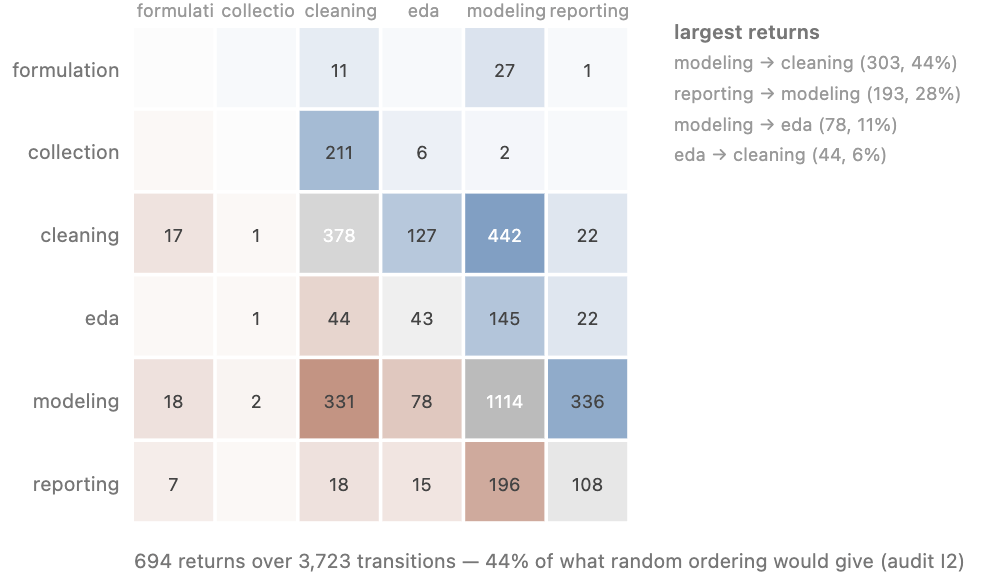}
  \caption{Transitions between DSLC stages within runs. Rows are the from-stage and columns are the to-stage; counts above the diagonal indicate forward progressions, while counts below the diagonal indicate returns to earlier stages.}
  \label{fig:dslc-turns}
\end{figure}

Forks in the charted garden were additionally aligned with the Data Science Life Cycle (DSLC) described in Veridical Data Science~\cite{yu2020veridical}. Figure~\ref{fig:dslc} illustrates how analyses distribute across DSLC stages. Figure~\ref{fig:dslc-turns} summarizes stage-to-stage movement within runs: analyses frequently iterate between data cleaning, exploratory data analysis (EDA), and modeling, and between reporting and modeling.

\subsection{Pivotal forks}
\label{sec:importance}

The raw option counts at a given fork conflate two cases: forks most analyses resolve the same way (with a few outliers), and forks analyses genuinely split on. We distinguish them with the \textit{effective number of options}: the inverse Herfindahl-Hirschman index (HHI) over option shares (including a null option, ``never visited'')~\cite{hirschman1945concentration,herfindahl1950concentration}. Using this measure, we define a fork as \textit{pivotal} if it meets both a coverage floor ($\geq$25\% of analyses reach it) and an effective-options floor ($>1.5$). \textit{17 forks} meet both criteria.

\begin{figure}
  \centering
  \includegraphics[width=\columnwidth,alt={Scatter plot of coverage versus effective number of options, highlighting forks that are both widely encountered and meaningfully contested (pivotal).}]{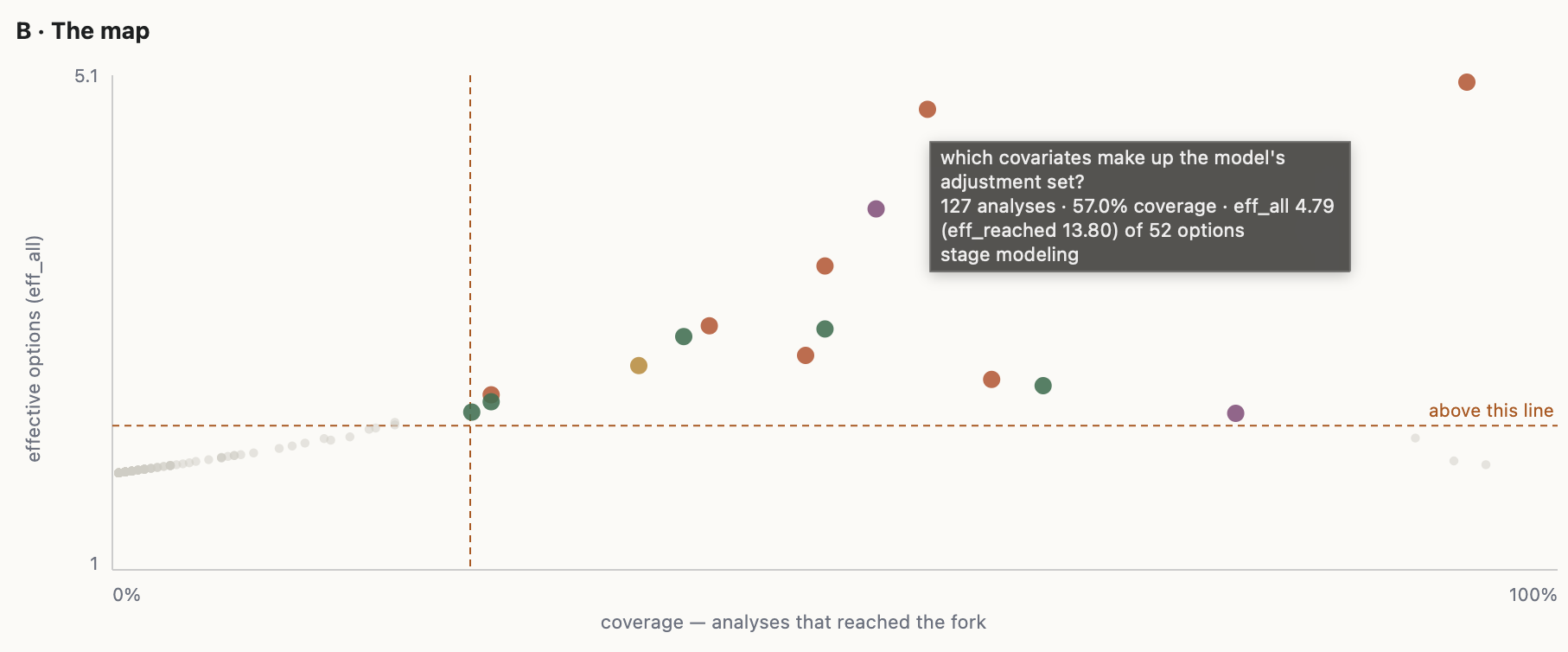}
  \caption{Coverage versus effective options for forks in the soccer case study. Upper-right indicates forks that are both widely encountered and meaningfully contested (\emph{pivotal}).}
  \label{fig:pivotalforks-atlas}
\end{figure}

 \textit{Do pivotal forks move the answer?} We attribute each run's reported outcome to a decision fork with a closed-form Shapley decomposition~\cite{shapley1953value} over a ridge surrogate fit on one-hot level indicators. Four payouts (outcomes) are evaluated: the binary verdict, $\log(\mathrm{OR})$, the $z$-statistic, and confidence-interval width. Attribution is reported per fork, aggregated over its options: it says where variation in the outcome concentrates, not which specific option or path would move an individual run's result, since that lookup is exactly what the viewer is designed not to encourage.

\begin{figure}
  \centering
  \includegraphics[width=\columnwidth,alt={Viewer screenshot showing Shapley attributions of pivotal forks to an outcome (verdict), with forks ranked or scored by attribution.}]{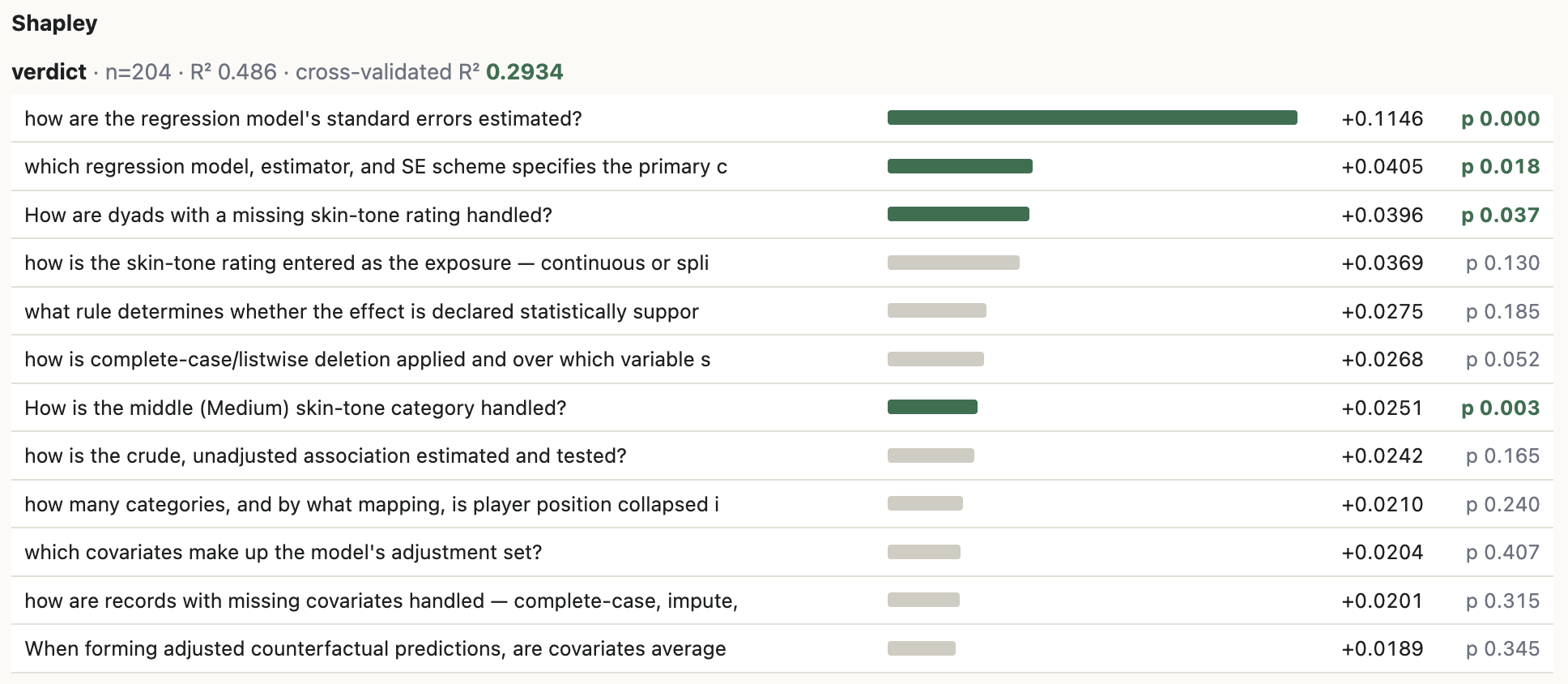}
  \caption{Which pivotal forks move the outcome. Screenshot from the viewer showing attributions for one outcome (verdict). Each fork is scored by its attribution to the reported result under the Shapley decomposition.}
  \label{fig:pivotalfork3}
\end{figure}

\textit{The expanded (contrast) set.} We compute attribution in the viewer over an \textit{expanded set of 36 forks}, not just the 17 pivotal ones: any fork with at least two levels each reached by $\geq$\MinRuns{} analyses, counting \emph{did not reach this fork} as a level. This targets a different question than the pivotal filter. Pivotal selects \emph{contested} forks (widely reached and genuinely splitting when reached); contrast selects forks whose signal can be \emph{attributed} at all, including rare, low-diversity forks that most analyses skip but that still cleanly separate those that do not (e.g., dropping a sensitive covariate, chosen by only 22 of 223 analyses). As shown in Table~\ref{tab:payoutcv} , expanding to 36 forks changes $R^2$ only modestly, indicating that the 17 pivotal forks dominate the predictive signal. Using only the pivotal set would exclude rare-but-consequential forks, so we use the expanded set as the main player set in the viewer.

\begin{table}
\centering
\caption{Out-of-sample prediction by payout, cross-validated $R^2$,
ridge penalty selected per payout by cross-validation over
$\{1,3,10,30,100,300,1000\}$ for both an expanded (contrast) set and the 17 pivotal forks.}
\label{tab:payoutcv}
\small
\begin{tabular}{@{}lrr@{}}
\toprule
\textbf{payout} & \textbf{$R^2$ (expanded, 36)} & \textbf{$R^2$ (pivotal, 17)} \\
\midrule
verdict          & $+0.293$ & $+0.262$ \\
$\log(\mathrm{OR})$ & $+0.081$ & $+0.032$ \\
$z$              & $+0.307$ & $+0.246$ \\
CI width         & $+0.038$ & $+0.046$ \\
\bottomrule
\end{tabular}
\end{table}

\section{Technical Evaluation}
\label{sec:techeval}

In this section, we evaluate whether the charting pipeline is \emph{stable} enough to support reliable analysis. Because \ForkSCOPE{} uses LLM prompts inside scripted stages, repeated executions can differ even on identical input. We therefore run the pipeline multiple times and report three complementary checks: (1) \emph{reproducibility of a build} via caching and provenance (where \emph{caching} means reusing stored LLM responses for identical calls, and \emph{provenance} means each result records the exact inputs/models/prompts that produced it); (2) \emph{stability under replication} across independent builds with an \emph{empty cache} (i.e., no stored LLM responses are reused, so all calls are re-made); and (3) a \emph{human benchmark} to contextualize how much variation is ``natural'' for the underlying clustering judgments. Full procedures and additional tables are in the supplement.

\subsection{Reproducibility (same build, same input)}
\label{sec:techeval-reproducibility}

\textit{Setup.} The model interface we use does not expose a controllable seed or temperature, so we do not rely on decoding parameters for exact replay. Instead, we cache every model call under a \emph{content-addressed} key (i.e., a hash of the prompt text, model identifier, input payload, and output schema), and we seed (or compute exactly) every non-model randomization step.

\textit{Results.} With caching enabled, re-running an unchanged stage is a \emph{cache hit} (the LLM call is not re-made) and returns byte-identical output. We also version every repaired vocabulary (never overwrite), and each reported number records the exact vocabulary tag that produced it, so figures are traceable and recomputable.

\textit{Observed hazards.} Auditing this mechanism surfaced two practical sources of non-reproducibility: (i) model \emph{aliases} can drift if they are repointed to newer versions, and (ii) a small amount of run metadata was initially recovered by re-reading an external corpus directory and could silently fail when that directory was missing. We mitigated both by recording resolved model versions at call time and treating failed metadata recovery as an error. Details are reported in the supplement.

\subsection{Stability under replication (independent builds)}
\label{sec:techeval-stability}

\textit{Setup.} We ran two full builds from the same corpus under isolated study roots (fresh output directories) with an empty cache, so every model call in the second build was a \emph{cache miss} and had to be re-made. We compare agreement across the pipeline's layers (decisions, options, and forks). To isolate the fork-induction stage from upstream variation, we additionally freeze two option-layer vocabularies and re-run fork induction five times against each (ten byte-identical-input replicates).

\textit{Results.} Fork induction shows variability with a stable core. Two independent builds recover the same high-level structure where our analyses rely on it: the well-populated \emph{trunk} converges exactly even when the rare tail does not. Every fork reached by more than \MinRuns{} runs in one build reproduces identically in the other (40 of 40), and so does every fork that clears the $\geq$25\% coverage bar (18 of 18)---the same coverage floor used to define \emph{pivotal} forks in \S\ref{sec:importance}, of which 17 of these 18 also clear that definition's effective-options half. Run-level \emph{geometry} (which forks a run touches) reproduces strongly (Mantel $r=0.766$~\cite{mantel1967detection}, i.e., a correlation between the two run-by-run ``which-forks-were-touched'' matrices, against a permutation null at zero). Substantive patterns also re-derive independently, including the persona-route association (\S\ref{sec:reading2}) and highway traffic rank ($\rho=0.951$). The remaining instability is concentrated in low-coverage forks and in forks whose disagreement is driven by rare options: small option-share fluctuations can tip whether a borderline option is merged, split, or treated as its own fork.

\begin{table}
\centering
\caption{Reproducibility by layer. Agreement falls with abstraction:
concrete operations reproduce most reliably, while induced forks are the most
variable layer.}
\label{tab:reliability}
\small
\begin{tabular}{@{}lll@{}}
\toprule
\textbf{layer} & \textbf{agreement} & \textbf{what it means} \\
\midrule
decisions & 0.713--0.895 & lenient--core $F_1$; median boundary jitter 1 line \\
options   & 0.853 & ARI (adjusted Rand index) over shared decisions \\
forks     & 0.63--0.77 & run-geometry corr.; membership agreement 0.47 \\
\bottomrule
\end{tabular}
\end{table}

\textit{Isolating fork induction.} Even holding the option layer fixed,
fork induction still varies: across ten replicate runs on byte-identical
input, it achieves ARI (adjusted Rand index)~\cite{hubert1985comparing} $=0.62$ with itself. Decomposing
this shortfall from perfect agreement, induction contributes the larger share
(58\%) relative to the inherited options layer. In other words, ``\Nforks{} forks'' is one
draw from a spread of roughly 281-317 on identical input, with most variation coming from the rarely visited tail rather than the high-coverage core.

\subsection{A human benchmark for ``natural'' variation}
\label{sec:techeval-humanbench}

\textit{Setup.} To interpret self-agreement, we benchmark against human
agreement on the same judgments. Three of the authors, working
independently of one another and blind to the pipeline's own verdicts,
labeled three tasks: (1) whether two code excerpts
implement the same decision (43 pairs), (2) whether two options are true
alternatives at the same fork (60 pairs), and (3) whether a sampled fork is
correctly formed and, if not, how it should be re-merged (30 forks).

\textit{Results.} At the option layer, pipeline agreement is within the range
of human agreement on the same items: two raters agree with each other 0.818
of the time, while two independent builds agree 0.849 of the time. When both
pipeline builds judged a pair as ``same option'' (19/19 pairs), the human
majority reached the same verdict ($p<0.0001$, exact binomial), suggesting
that agreement at this layer reflects correct canonicalization rather than
shared error.

The benchmark also surfaced an actionable limitation: raters repeatedly
identified a small set of induced forks as composite, conflating
\emph{which variables enter the model} with \emph{how each variable is
transformed}, suggesting a future skill that allows revising forks based 
on human inputs. Full per-task results and limitations appear in the supplement.

\subsection{Computation}
\label{sec:computation}
We ran the pipeline using a Claude Max subscription; the figures below are
the \emph{reported} per-call costs logged by the API.
Charting the pooled 223-analysis vocabulary cost \$804 in logged API spend
across 2{,}273 model calls: distillation (extracting decisions from code and
prose) accounts for \$447 (680 calls, \$1.91/analysis), vocabulary clustering
\$112 (277 calls), the repair chain \$130 (1{,}154 calls), and the
code-ablation audit \$114 (162 calls). Two models are used by role: 
Sonnet performs bulk extraction (724 calls, \$444 total,
\$0.61/call) and Opus performs adjudication: the shorter same/different
judgments (1{,}569 calls, \$362 total, \$0.21/call). The ten frozen-option
fork-formation replicates used for the fork-induction stability check
(\S\ref{sec:techeval-stability}) cost roughly \$36 each (\$356 total).

\section{Discussion}
\label{sec:discussion}

\subsection{Design implications for AI-analyst tooling}

Our comparison against Miao et al.~\cite{miao2026agentic} and Bertran et al.~\cite{bertran2026many} (\S\ref{sec:aiagents}) suggests three design choices for building the next agentic multiverse corpus, and for using the garden it produces, independent of whether \ForkScope{} is adopted.

First, the generation interface can either enumerate a fixed decision space upfront or leave the space open and chart it afterward. When the downstream question is focused, a fixed grid is practical. In our corpus, many high-coverage forks have no slot in an axis list written in advance. A direct comparison against this corpus's curated codebook quantifies the gap: our induced vocabulary recovers at least 91\% of Bertran et al.\ (2026)\cite{bertran2026many}'s eleven curated decision points , while 60\% of our most-walked decision points, including several reached by over half the corpus, have no counterpart in that codebook. In practice, open-ended generation surfaces choices such as specification search outside delivered code or globally suppressed warnings.

Second, auditing can operate at the level of whole analyses or at the level of options within forks. Whole-run verdicts are efficient but coarse: one flawed step can fail an otherwise sound run. Option-level auditing keeps uncontested parts of a run usable for the purpose of charting the garden when only one choice is in question and produces concrete exhibits tied to a specific fork and option. 

Third, a charted garden is not only a target for human review; it is scaffolding a person can walk on their own terms. Once decisions are organized into forks with audited options, a garden visitor can specify an \textit{itinerary}, i.e., the option their prior beliefs find defensible at each contested fork, and have \ForkSCOPE{} check it for internal incompatibility before comparing it against nearby analyses (\S\ref{sec:skills}, \texttt{/itinerary}). This workflow uses AI scale to chart the space and human judgment to walk it, yielding a scoped account of uncertainty without generating and auditing a full multiverse, and helping keep open-ended multiverses from collapsing into an analytic black hole~\cite{delgiudice2021traveler}. Because the viewer does not reveal which option would move an individual outcome (\S\ref{sec:viewer}), itinerary-walking asks visitors to commit to options on their own merits rather than searching for a path that flatters a preferred conclusion. 

The first two choices carry cost: open-ended generation increases charting effort, and option-level auditing increases per-decision review. We argue that both are worth paying once analysis generation becomes cheap enough that the bottleneck shifts to interpretation and verification. The third has not yet been evaluated with real garden visitors; \S\ref{sec:humanvsai}'s result that the human and agentic corpora largely agree on the shape of the decision space is a first step toward showing the garden is worth a human visiting. \ForkScope{} is intended to reduce the heavier workload of charting and facilitate easier review by humans.

\subsection{Limitations and future work}
\label{sec:limitations}
We treat the following limitations as open design challenges and research questions. \ForkScope{} offers a \emph{ground-up} way to operationalize and navigate an agentic multiverse, but a substantial need for further research remains before it can support rigorous end-to-end multiverse inference.

\textit{Scope.} Our evidence is a single in-depth case study used to instantiate the framework and surface concrete failure modes. Transfer to other domains is untested. The corpus reflects a single generation design, so coverage of the full analytic space is unknown. Whether the agent multiverse induced here overlaps with the human one remains an open question noted by the corpus authors~\cite{bertran2026many}.

\textit{Attribution.} Our fork-to-outcome attribution
(\S\ref{sec:importance}) is observational. Choices co-occur non-randomly and
were not experimentally assigned. We compute attribution
on a surrogate fit to reported outcomes, so scores reflect both confounding
and surrogate approximation error.

\textit{LLM dependence.} We rely on Claude models both to implement
parts of the pipeline and to run it (\S\ref{sec:architecture}), and the
resulting chart is sensitive to that choice. At runtime, open-weight and
smaller code-specialized models could plausibly substitute for more
mechanical parsing stages and reduce the cost reported in \S\ref{sec:computation}.
The silent-decision rate reported in \S\ref{sec:corpus} is also model-dependent
because it is recovered from code-prose linking. In one comparison,
switching the reader model from Opus to Sonnet roughly doubled the observed
rate. We have not yet validated which reading is closer to correct, and for
this task validation is limited to annotator agreement.

\textit{Atomicity of extracted actions.} We treat each extracted action from code
as one raw decision point. The human benchmark (\S\ref{sec:techeval-humanbench})
shows that some induced forks bundle multiple independent choices (e.g., which
variable enters a model and how it is transformed). Automatic detection and
decomposition of composite decisions is future work.

\textit{No user study yet.} This paper reports the framework design and
technical evaluation of the chart it produces. We have not yet studied how
garden owners prefer to provide corrections on contested forks. As an initial,
systematic proxy, part of \S\ref{sec:techeval} uses a simulated AI garden owner
to probe review decisions more broadly than one person's schedule allows. We
plan to scale this protocol alongside studies with human garden owners.

\textit{Cross-corpus comparisons.} We pool the human and agent runs into a
single vocabulary rather than charting them separately: all \Nforks{} forks
and \Nfamilies{} options are shared slots that both corpora occupy, and the
viewer’s corpus filter (\S\ref{sec:skills}) allows either subset to be
examined within the same map. However, the two corpora are recovered with
different reliability (\S\ref{sec:corpus}), so we do not interpret the data
as supporting a clean outcome-level human-AI comparison. The structural
comparison in \S\ref{sec:humanvsai} is correspondingly narrower: it provides
evidence of rough compatibility between the induced vocabularies, but the
human subset is too small and inconsistently recovered to support claims
about which decisions each corpus favors.

\textit{Conservative canonicalization.} We prefer under-merging to
over-merging during canonicalization. This makes spurious consensus less
likely, but it also means \Nforks{} and downstream counts are likely 
overestimates of the complexity of the charted garden. The size of the
stable core (\S~\ref{sec:techeval-stability}) is a more reliable indicator.

\section{Conclusion}
\label{sec:conclusion}

We presented \ForkSCOPE{}, a bottom-up human-AI collaboration framework for charting and reading an agentic analysis corpus without a taxonomy fixed before or after generation. Applied to a pooled corpus of 223 agent- and human-written analyses of the same many-analyst question, \ForkSCOPE{} induced \Nforks{} forks over \Nfamilies{} options, distinguished widely contested \emph{pivotal} forks from settled conventions, and showed that a modest subset of forks carries much of the out-of-sample predictive signal for reported outcomes. A three-rater benchmark found the pipeline's self-agreement at the option layer comparable to independent human agreement on the same items, while also surfacing concrete directions for future work (\S\ref{sec:limitations}).

\ForkSCOPE{} is designed to keep human oversight tractable as agentic corpora grow past what any one reviewer can read end-to-end. It combines a reproducible, artifact-based instrument (pinned prompts, cached builds, and explicit review gates) with an interactive viewer that supports systematic inspection of forks and options (e.g., tags, associations, pivotal-fork impact views, lifecycle dynamics, and run-level browsing). Because the induced map follows the fork/option abstraction expected by existing multiverse tools (\S\ref{sec:datamodel}), a charted garden can also serve as a starting point for downstream multiverse visualization and analysis.

\bibliographystyle{plainnat}
\bibliography{refs}

@misc{openai_chatgpt,
  author       = {{OpenAI}},
  title        = {ChatGPT},
  howpublished = {Large language model},
  year         = {2026},
  note         = {Accessed 2026-09-09}
}

@article{hirschman1945concentration,
  author  = {Hirschman, Albert O.},
  title   = {National Power and the Structure of Foreign Trade},
  journal = {University of California Press},
  year    = {1945}
}

@article{herfindahl1950concentration,
  author  = {Herfindahl, Orris C.},
  title   = {Concentration in the U.S. Steel Industry},
  journal = {Columbia University},
  year    = {1950},
  note    = {Unpublished doctoral dissertation}
}

@article{shapley1953value,
  author  = {Shapley, Lloyd S.},
  title   = {A Value for $n$-Person Games},
  journal = {Contributions to the Theory of Games},
  year    = {1953},
  editor  = {Kuhn, Harold W. and Tucker, Albert W.},
  publisher = {Princeton University Press},
  pages   = {307--317}
}

@article{mantel1967detection,
  author  = {Mantel, Nathan},
  title   = {The Detection of Disease Clustering and a Generalized Regression Approach},
  journal = {Cancer Research},
  year    = {1967},
  volume  = {27},
  number  = {2},
  pages   = {209--220}
}

@article{hubert1985comparing,
  author  = {Hubert, Lawrence and Arabie, Phipps},
  title   = {Comparing Partitions},
  journal = {Journal of Classification},
  year    = {1985},
  volume  = {2},
  number  = {1},
  pages   = {193--218}
}

@unpublished{gelman2013garden,
  author = {Gelman, Andrew and Loken, Eric},
  title  = {The Garden of Forking Paths: Why Multiple Comparisons Can Be a
            Problem, Even When There Is No ``Fishing Expedition'' or
            ``p-Hacking'' and the Research Hypothesis Was Posited Ahead of Time},
  note   = {Unpublished manuscript, Department of Statistics,
            Columbia University},
  year   = {2013},
  url    = {http://stat.columbia.edu/~gelman/research/unpublished/forking.pdf}
}

@article{gelman2014statistical,
  author  = {Gelman, Andrew and Loken, Eric},
  title   = {The Statistical Crisis in Science},
  journal = {American Scientist},
  volume  = {102},
  number  = {6},
  pages   = {460--465},
  year    = {2014}
}

@article{simmons2011false,
  author  = {Simmons, Joseph P. and Nelson, Leif D. and Simonsohn, Uri},
  title   = {False-Positive Psychology: Undisclosed Flexibility in Data
             Collection and Analysis Allows Presenting Anything as Significant},
  journal = {Psychological Science},
  volume  = {22},
  number  = {11},
  pages   = {1359--1366},
  year    = {2011},
  doi     = {10.1177/0956797611417632}
}

@article{steegen2016multiverse,
  author  = {Steegen, Sara and Tuerlinckx, Francis and Gelman, Andrew and
             Vanpaemel, Wolf},
  title   = {Increasing Transparency Through a Multiverse Analysis},
  journal = {Perspectives on Psychological Science},
  volume  = {11},
  number  = {5},
  pages   = {702--712},
  year    = {2016},
  doi     = {10.1177/1745691616658637}
}

@article{simonsohn2020specification,
  author  = {Simonsohn, Uri and Simmons, Joseph P. and Nelson, Leif D.},
  title   = {Specification Curve Analysis},
  journal = {Nature Human Behaviour},
  volume  = {4},
  number  = {11},
  pages   = {1208--1214},
  year    = {2020},
  doi     = {10.1038/s41562-020-0912-z}
}

@article{silberzahn2018many,
  author  = {Silberzahn, Raphael and Uhlmann, Eric L. and Martin, Daniel P.
             and others},
  title   = {Many Analysts, One Data Set: Making Transparent How Variations
             in Analytic Choices Affect Results},
  journal = {Advances in Methods and Practices in Psychological Science},
  volume  = {1},
  number  = {3},
  pages   = {337--356},
  year    = {2018},
  doi     = {10.1177/2515245917747646}
}

@article{botviniknezer2020variability,
  author  = {Botvinik-Nezer, Rotem and Holzmeister, Felix and
             Camerer, Colin F. and others},
  title   = {Variability in the Analysis of a Single Neuroimaging Dataset by
             Many Teams},
  journal = {Nature},
  volume  = {582},
  number  = {7810},
  pages   = {84--88},
  year    = {2020},
  doi     = {10.1038/s41586-020-2314-9}
}

@article{breznau2022observing,
  author  = {Breznau, Nate and Rinke, Eike Mark and Wuttke, Alexander and
             others},
  title   = {Observing Many Researchers Using the Same Data and Hypothesis
             Reveals a Hidden Universe of Uncertainty},
  journal = {Proceedings of the National Academy of Sciences},
  volume  = {119},
  number  = {44},
  pages   = {e2203150119},
  year    = {2022},
  doi     = {10.1073/pnas.2203150119}
}

@inproceedings{dragicevic2019increasing,
  author    = {Dragicevic, Pierre and Jansen, Yvonne and Sarma, Abhraneel and
               Kay, Matthew and Chevalier, Fanny},
  title     = {Increasing the Transparency of Research Papers with Explorable
               Multiverse Analyses},
  booktitle = {Proceedings of the 2019 CHI Conference on Human Factors in
               Computing Systems},
  year      = {2019},
  doi       = {10.1145/3290605.3300295}
}

@article{liu2021boba,
  author  = {Liu, Yang and Kale, Alex and Althoff, Tim and Heer, Jeffrey},
  title   = {Boba: Authoring and Visualizing Multiverse Analyses},
  journal = {IEEE Transactions on Visualization and Computer Graphics},
  volume  = {27},
  number  = {2},
  pages   = {1753--1763},
  year    = {2021},
  doi     = {10.1109/TVCG.2020.3028985}
}

@inproceedings{sarma2023multiverse,
  author    = {Sarma, Abhraneel and Kale, Alex and Moon, Michael and
               Taback, Nathan and Chevalier, Fanny and Hullman, Jessica and
               Kay, Matthew},
  title     = {multiverse: Multiplexing Alternative Data Analyses in R
               Notebooks},
  booktitle = {Proceedings of the 2023 CHI Conference on Human Factors in
               Computing Systems},
  year      = {2023},
  doi       = {10.1145/3544548.3580726}
}

@inproceedings{sarma2024milliways,
  author    = {Sarma, Abhraneel and Hwang, Kyle and Hullman, Jessica and
               Kay, Matthew},
  title     = {Milliways: Taming Multiverses through Principled Evaluation of
               Data Analysis Paths},
  booktitle = {Proceedings of the 2024 CHI Conference on Human Factors in
               Computing Systems},
  year      = {2024},
  doi       = {10.1145/3613904.3642375}
}

@article{hall2022survey,
  author  = {Hall, Brian D. and Liu, Yang and Jansen, Yvonne and
             Dragicevic, Pierre and Chevalier, Fanny and Kay, Matthew},
  title   = {A Survey of Tasks and Visualizations in Multiverse Analysis
             Reports},
  journal = {Computer Graphics Forum},
  volume  = {41},
  number  = {1},
  pages   = {402--426},
  year    = {2022},
  doi     = {10.1111/cgf.14443}
}

@inproceedings{liu2020paths,
  author    = {Liu, Yang and Althoff, Tim and Heer, Jeffrey},
  title     = {Paths Explored, Paths Omitted, Paths Obscured: Decision Points
               and Selective Reporting in End-to-End Data Analysis},
  booktitle = {Proceedings of the 2020 CHI Conference on Human Factors in
               Computing Systems},
  year      = {2020},
  doi       = {10.1145/3313831.3376533}
}

@article{bertran2026many,
  author  = {Bertran, Martin and Fogliato, Riccardo and Wu, Zhiwei Steven},
  title   = {Many {AI} Analysts, One Dataset: Navigating the Agentic Data Science Multiverse},
  journal = {Proceedings of the National Academy of Sciences of the United States of America},
  year    = {2026},
  volume  = {123},
  number  = {29},
  pages   = {e2606495123},
  month   = jul,
  doi     = {10.1073/pnas.2606495123}
}

@article{miao2026agentic,
  author  = {Miao, Jiacheng and Pritchard, Jonathan K. and Zou, James},
  title   = {The Agentic Garden of Forking Paths},
  journal = {arXiv preprint arXiv:2607.01507},
  year    = {2026},
  doi     = {10.48550/arXiv.2607.01507}
}

@article{asher2026phack,
  author  = {Asher, Samuel G. Z. and Malzahn, Janet and Persano, Jessica M.
             and Paschal, Elliot J. and Myers, Andrew C. W. and
             Hall, Andrew B.},
  title   = {Do Claude Code and Codex p-Hack? Sycophancy and Statistical
             Analysis in Large Language Models},
  year    = {2026}
}

@software{anthropic2025claudecode,
  author = {{Anthropic}},
  title  = {Claude Code (Version 2.1) [Computer software]},
  year   = {2025},
  url    = {https://code.claude.com}
}

@inproceedings{gu2024wizard,
  author    = {Gu, Ken and Grunde-McLaughlin, Madeleine and McNutt, Andrew
               and Heer, Jeffrey and Althoff, Tim},
  title     = {How Do Data Analysts Respond to AI Assistance? A Wizard-of-Oz
               Study},
  booktitle = {Proceedings of the 2024 CHI Conference on Human Factors in
               Computing Systems},
  year      = {2024},
  doi       = {10.1145/3613904.3641891}
}

@inproceedings{gu2024verify,
  author    = {Gu, Ken and Shang, Ruoxi and Althoff, Tim and Wang, Chenglong
               and Drucker, Steven M.},
  title     = {How Do Analysts Understand and Verify AI-Assisted Data
               Analyses?},
  booktitle = {Proceedings of the 2024 CHI Conference on Human Factors in
               Computing Systems},
  year      = {2024},
  doi       = {10.1145/3613904.3642497}
}

@article{rewolinski2026sanity,
  author  = {Rewolinski, Zachary T. and Zane, Austin V. and Huang, Hao and
             Singh, Chandan and Wang, Chenglong and Gao, Jianfeng and
             Yu, Bin},
  title   = {Sanity Checks for Agentic Data Science},
  journal = {arXiv preprint arXiv:2604.11003},
  year    = {2026}
}

@article{messeri2024artificial,
  author  = {Messeri, Lisa and Crockett, M. J.},
  title   = {Artificial Intelligence and Illusions of Understanding in
             Scientific Research},
  journal = {Nature},
  volume  = {627},
  pages   = {49--58},
  year    = {2024},
  doi     = {10.1038/s41586-024-07146-0}
}

@article{yu2020veridical,
  author  = {Yu, Bin and Kumbier, Karl},
  title   = {Veridical Data Science},
  journal = {Proceedings of the National Academy of Sciences},
  volume  = {117},
  number  = {8},
  pages   = {3920--3929},
  year    = {2020},
  doi     = {10.1073/pnas.1901326117}
}

@article{delgiudice2021traveler,
  author  = {{Del Giudice}, Marco and Gangestad, Steven W.},
  title   = {A Traveler's Guide to the Multiverse: Promises, Pitfalls, and
             a Framework for the Evaluation of Analytic Decisions},
  journal = {Advances in Methods and Practices in Psychological Science},
  volume  = {4},
  number  = {1},
  pages   = {1--15},
  year    = {2021},
  doi     = {10.1177/2515245920954925}
}

@article{heyman2025crowdsourcing,
  author  = {Heyman, Tom and Pronizius, Ekaterina and Lewis, Savannah C.
             and others},
  title   = {Crowdsourcing Multiverse Analyses to Explore the Impact of
             Different Data-Processing and Analysis Decisions: A Tutorial},
  journal = {Psychological Methods},
  year    = {2025},
  doi     = {10.1037/met0000770}
}

@inproceedings{kale2019decision,
  author    = {Kale, Alex and Kay, Matthew and Hullman, Jessica},
  title     = {Decision-Making Under Uncertainty in Research Synthesis:
               Designing for the Garden of Forking Paths},
  booktitle = {Proceedings of the 2019 CHI Conference on Human Factors in
               Computing Systems},
  year      = {2019},
  doi       = {10.1145/3290605.3300432}
}

@inproceedings{merrill2021multiverse,
  author    = {Merrill, Mike A. and Zhang, Ge and Althoff, Tim},
  title     = {MULTIVERSE: Mining Collective Data Science Knowledge from
               Code on the Web to Suggest Alternative Analysis Approaches},
  booktitle = {Proceedings of the 27th ACM SIGKDD Conference on Knowledge
               Discovery \& Data Mining},
  year      = {2021},
  pages     = {1212--1222},
  doi       = {10.1145/3447548.3467455}
}

@book{yubarter2024veridical,
  author    = {Yu, Bin and Barter, Rebecca L.},
  title     = {Veridical Data Science: The Practice of Responsible Data
               Analysis and Decision Making},
  publisher = {MIT Press},
  year      = {2024}
}

@article{gomez2025taxonomy,
  author  = {Gomez, Catalina and Cho, Sue Min and Ke, Shichang and
             Huang, Chien-Ming and Unberath, Mathias},
  title   = {Human-{AI} Collaboration Is Not Very Collaborative yet: A
             Taxonomy of Interaction Patterns in {AI}-Assisted Decision
             Making from a Systematic Review},
  journal = {Frontiers in Computer Science},
  volume  = {6},
  year    = {2025},
  doi     = {10.3389/fcomp.2024.1521066}
}

@misc{lou2025teaming,
  author  = {Lou, Bowen and Lu, Tian and Raghu, T. S. and Zhang, Yingjie},
  title   = {Unraveling Human-{AI} Teaming: A Review and Outlook},
  year    = {2025},
  eprint  = {2504.05755},
  archiveprefix = {arXiv},
  doi     = {10.48550/arXiv.2504.05755}
}

@article{benmichael2025doesai,
  author  = {{Ben-Michael}, Eli and Greiner, D. James and Huang, Melody
             and Imai, Kosuke and Jiang, Zhichao and Shin, Sooahn},
  title   = {Does {AI} Help Humans Make Better Decisions? A Statistical
             Evaluation Framework for Experimental and Observational
             Studies},
  journal = {Proceedings of the National Academy of Sciences},
  volume  = {122},
  number  = {38},
  pages   = {e2505106122},
  year    = {2025},
  doi     = {10.1073/pnas.2505106122}
}

@inproceedings{liao2020questioning,
  author    = {Liao, Q. Vera and Gruen, Daniel and Miller, Sarah},
  title     = {Questioning the {AI}: Informing Design Practices for
               Explainable {AI} User Experiences},
  booktitle = {Proceedings of the 2020 CHI Conference on Human Factors in
               Computing Systems},
  year      = {2020},
  doi       = {10.1145/3313831.3376590}
}

@inproceedings{kazemitabaar2024steering,
  author    = {Kazemitabaar, Majeed and Williams, Jack and Drosos, Ian
               and Grossman, Tovi and Henley, Austin Zachary and
               Negreanu, Carina and Sarkar, Advait},
  title     = {Improving Steering and Verification in {AI}-Assisted Data
               Analysis with Interactive Task Decomposition},
  booktitle = {Proceedings of the 37th Annual ACM Symposium on User
               Interface Software and Technology},
  year      = {2024},
  doi       = {10.1145/3654777.3676345}
}

@article{narayanan2018humans,
  author  = {Narayanan, Menaka and Chen, Emily and He, Jeffrey and
             Kim, Been and Gershman, Samuel J. and Doshi-Velez, Finale},
  title   = {How Do Humans Understand Explanations from Machine Learning
             Systems? An Evaluation of the Human-Interpretability of
             Explanation},
  journal = {arXiv preprint arXiv:1802.00682},
  year    = {2018},
  doi     = {10.48550/arXiv.1802.00682}
}

@article{lai2023selective,
  author  = {Lai, Vivian and Zhang, Yiming and Chen, Chacha and
             Liao, Q. Vera and Tan, Chenhao},
  title   = {Selective Explanations: Leveraging Human Input to Align
             Explainable {AI}},
  journal = {Proceedings of the ACM on Human-Computer Interaction},
  volume  = {7},
  number  = {CSCW2},
  year    = {2023},
  doi     = {10.1145/3610206}
}

@article{zhao2025lightva,
  author  = {Zhao, Yuheng and Wang, Junjie and Xiang, Linbing and
             Zhang, Xiaowen and Guo, Zifei and Turkay, Cagatay and
             Zhang, Yu and Chen, Siming},
  title   = {{LightVA}: Lightweight Visual Analytics with {LLM}
             Agent-Based Task Planning and Execution},
  journal = {IEEE Transactions on Visualization and Computer Graphics},
  volume  = {31},
  number  = {9},
  pages   = {6162--6177},
  year    = {2025},
  doi     = {10.1109/TVCG.2024.3496112}
}

\end{document}